\documentclass[aps, prfluids, amsmath, amssymb, longbibliography]{revtex4-2}
\usepackage{epstopdf, epsfig}
\usepackage{mathrsfs}
\usepackage{amsmath}
\usepackage{graphicx}
\usepackage{tikz}
\usepackage{bm}
\usepackage{xcolor}
\usepackage[colorlinks = true,
            linkcolor = blue,
            urlcolor  = blue,
            citecolor = blue,
            anchorcolor = blue]{hyperref}

\newcommand{\dd}{\mathrm{d}}
\renewcommand{\vec}[1]{\mathbf{#1}}
\newcommand{\bhat}[1]{\hat{\mathbf{#1}}}

\begin{document}

  \title{Closely spaced co-rotating helical vortices: General solutions}
  \author{A. Castillo-Castellanos}
    \affiliation{Aix Marseille Universit\'e, CNRS, Centrale Marseille, IRPHE, Marseille, France}
    \email{andres-alonso.castillo-castellanos@univ-amu.fr}
  \author{E. Dur\'an Venegas}%
    \affiliation{Aix Marseille Universit\'e, CNRS, Centrale Marseille, IRPHE, Marseille, France}
    \affiliation{Departamento de Bioingenier\'ia e Ingenier\'ia Aeroespacial, Universidad Carlos III de Madrid,
    Spain}
  \author{S. Le Diz\`es}
    \affiliation{Aix Marseille Universit\'e, CNRS, Centrale Marseille, IRPHE, Marseille, France}
  \date{\today}

	\pacs{~}
	\keywords{}

  \begin{abstract}
    In this work, we present general solutions for closely spaced co-rotating
    helical vortices using a filament approach. For these vortex structures,
    helical symmetry is broken, but solutions maintain a form of spatial
    periodicity. We show there exists a moving frame where vortex elements move
    along the structure without distorting it. These solutions can be used to
    characterize the evolution of the twin-tip vortices produced in the wake of
    a tip-splitting rotor blade. The resulting wake has a dual nature.
    Locally, the structure behaves much like an helical pair aligned with the
    locally tangent flow. However, as we move away from the vortex structure the
    induced flow is reminiscent of an `equivalent' helical vortex with thicker
    core.
  \end{abstract}

  \maketitle

  \section{Introduction}
  \label{sec1}

  Rotating blades, such as those of a helicopter rotor or a horizontal-axis wind
  turbine, generate concentrated vortices at their tips, transported downstream,
  creating a persistent helical pattern. These concentrated vortices can have a
  core size as small as 1\% of the rotor radius \cite{young2003vortex}. The
  structure and stability of these helical vortices are associated with several
  practical issues actively investigated today. One of these issues concerns the
  interaction between a tip-vortex and a downstream surface (Vortex-Surface
  Interactions), which causes significant noise and vibration problems. In
  particular, the interaction between a tip-vortex and a rotor blade
  (Blade-Vortex Interaction, BVI) is responsible for reduced power output and
  premature structural component fatigue in wind turbines
  \cite{hansen2012impact, barthelmie2010evaluation}, and undesirable noise
  in helicopter rotors during low speed and descending flight
  \citep{gandhi2000influence}. BVI mitigation methods can be classified as
  active and passive \cite{hardin1987concepts,YU200097, brocklehurst2013review}.
  Active methods include dynamical modifications of the wake, for instance, by
  steering the wake towards a particular direction \cite{fleming2017field,
  campagnolo2020wind}, or by introducing a periodic perturbation
  \cite{quaranta2015long, quaranta2019local, huang2019numerical}. Passive
  methods use a modified blade geometry to enhance particular flow features, for
  instance, by introducing serrations on the leading edge
  \cite{ito2009,pang2018investigation} or through the use of winglets
  \cite{chattot2009effects,ebrahimi2018tip} and slotted tips
  \cite{han2004investigation}. One alternative proposed by
  \citet{brocklehurst1994reduction} is to use a modified airfoil to split the
  tip vortex into two closely spaced vortices. As the tip vortices are advected
  downstream, they interact with each other and eventually merge back into a
  large more diffuse vortex \cite{Schroder2021}. This approach offers
  significant noise reduction and has no adverse effects on control load or
  performance \citep{copland1998experimental, coton2005analysis,
  Cho2006NumericalIO,Jung2008EffectOT}, but little is known about the wake
  structure and the mechanism leading to a larger vortex. One of our motivations
  is to provide information on the vortical wake that could exist before
  the merging process. We consider an ideal framework of vortex filaments to
  illustrate the type of structures that can be created by the interaction
  of two closely spaced but distinct helical vortices.

  A wide array of models, including analytical, experimental, and numerical
  approaches for rotor wake aerodynamics are available
  \cite{sanderse2011review}. Direct numerical simulations, which require solving
  the Navier-Stokes equations for all scales of fluid motion up to the
  Kolmogorov scale remain prohibitively expensive for realistic configurations.
  These numerical simulations require a turbulence model, such as the
  Reynolds-averaged Navier-Stokes (RANS) or large eddy simulation (LES), to take
  into account the effect of the unresolved small scales
  \cite{sanderse2011review, stevens2017flow, caprace2020wakes,
  branlard2017wind}. These simulations may take into account the rotor directly,
  by solving the boundary layer along the rotor blade, or indirectly, by
  representing the rotor blades as body forces  as in the actuator line method
  (ACL). For applications, the design and optimization of rotors often rely on
  computationally inexpensive models that describe the aerodynamic loads and
  wake dynamics. In the wind turbine community, blade loading is usually
  obtained using the Blade Element Momentum (BEM) theory, which is based on the
  actuator disc principle, corresponding to an infinitesimally thin rotor with
  an infinite number of blades. This approach has been progressively extended to
  include a finite number of blades, three-dimensional effects, unsteady flow
  loading \cite{sorensen2016general}, and  blade thickness \cite{van2015rotor}.
  But using BEM requires a wake model and airfoil data
  \cite{shen2009determination,sorensen2011aerodynamic, sorensen2016general}.
  Wakes with a prescribed helical symmetry are often used
  \cite{goldstein1929vortex,okulov2007stability}. Such wakes do not account for
  the inboard motion and subsequent radial contraction (resp. expansion)
  observed in helicopter flight (resp. wind turbine) regimes.
  Additionally, a prescribed helical wake simply cannot describe the tip-vortex
  interaction considered here.

  A free vortex method is more adequate to describe the structure and dynamics
  of the wake (see for instance \cite{leishman2002free}). Although the inboard
  motion associated with vortex formation is not easily described, the
  free-vortex method captures the contraction/expansion of the wake. This method
  has been used to extend the analytical solution of a single infinite helix
  \cite{kawada1936induced, hardin1982velocity} to multiple vortices
  \cite{venegas2019generalized} and semi-infinite configurations
  \cite{miller1993direct,venegas2020}.
  Using Joukowski's wake model, for which there is only one free tip-vortex per
  blade, \citet{venegas2020} were able to find  steady solutions describing the
  wake in all the vertical flight regimes of a helicopter (hover, climb,
  descent) and all wind turbine regimes, including cases for which the vortex
  structure is strongly deformed in the near wake and crosses the rotor plane.

  For Joukowski's model, the solution matches in the far-field a uniform helix
  solution. When two (or more) free vortices are emitted by each blade, the
  solution becomes more complex even in the far-field. As shown by
  \cite{venegas2019generalized} for two counter-rotating vortices, steady
  solutions can still be obtained. These solutions are no longer uniform but
  still exhibit a spatial periodicity property. In the present work, we shall use the method
  developed in \cite{venegas2020,venegas2019generalized} to describe the wake of
  tip-splitting rotors.
%

  As for counter-rotating vortices, we expect contraction/expansion of the wake
  close to the rotor as well as a complex far-field. The closest canonical
  example to these closely spaced co-rotating helical vortex pairs is a pair of
  equistrength parallel vortices. In such a case, the parameters driving the
  vortex dynamics are the circulation $\Gamma$, separation distance $d$, and
  vortex core size $a$. In general, $a$ grows with time due to viscous diffusion
  of vorticity with $a(t)=\sqrt{4\nu t + a_0^2}$ where $\nu$ is the kinematic
  viscosity, while $d$ remains virtually constant. Simultaneously, the vortex
  pair rotates around the total vortex centroid with a constant rotation rate
  $\omega=\Gamma/(\pi d^2)$ \cite{ledizes2002viscous}. When the ratio $a/d$
  exceeds a threshold value around 0.23, $d$ suddenly decreases
  and the vortices merge into a single entity (see, for instance
  \cite{melander1988symmetric, meunier2002merging, cerretelli2003physical,
  josserand2007merging, leweke2016dynamics}).
  In the present work, we neglect viscous diffusion and consider a sufficiently
  small value of $a/d$, typically 0.1 such that the vortices remain distinct and
  the filament approach justified.  We shall see that the rotation motion of the
  two vortices around each other is also present in our solutions.


  The paper is organized as follows. Section \ref{sec:2} briefly introduces the
  vortex filament framework used. Section \ref{sec:3} is concerned with the
  far-field. We show how steady spatially-periodic solutions are obtained from a
  few geometric parameters. In section \ref{sec4},  these solutions are then
  used to compute the far-field contribution for the near-field solutions
  obtained close to a tip-splitting rotor. We present solutions representative
  of both wind turbine and helicopter regimes and discuss the rotor influence on
  the wake structure. We also provide a simple model explaining the observed
  spatial variations of the parameters.


  \section{Framework}
  \label{sec:2}

  Our problem is defined by three spatial scales. The radius $R$ of the rotor,
  the separation distance $d$, and the vortex core size $a$.
  The filament approach is perfectly justified as long as the
  core size remains small compared to other spatial scales
  (separation distances, local curvature radius).
  Closely spaced filament vortices correspond to situations for which
  $a\ll d \ll R$. We use the same framework as in
  \cite{venegas2019generalized, venegas2020}. All the vorticity is considered as
  being concentrated along lines which move as material lines in the fluid
  according to
  \begin{equation}\label{2.1:time}
    \frac{\dd \vec{X}_i}{\dd t} = \vec{U}(\vec{X}_i) + \vec{U}^{\infty}
  \end{equation}
  where $\vec{X}_i$ is the position vector of the $i$-th vortex filament,
  $\vec{U}$ the velocity induced by the vortices and $\vec{U}^\infty$ an
  external velocity field. The induced velocity $\vec{U}$ is given by the
  Biot-Savart law
  \begin{equation}\label{2.2:bs}
    \vec{U} (\vec{X}_i) =
    \sum_{j=1}^{n} \frac{\Gamma_j}{4\pi}
    \int \frac{(\vec{X}_j-\vec{X}_i) \times \mathrm{d}\vec{T}_j}{\vert(\vec{X}_j-\vec{X}_i)\vert^2}
  \end{equation}
  where the integrals cover each vortex filament defined by its circulation
  $\Gamma_j$, position vector $\vec{X}_j$ and corresponding tangent unit vector
  $\vec{T}_j$. Filaments are discretised in straight segments
  $[\vec{X}_i^n, \vec{X}_i^{n+1}]$ in order to compute the velocity field and
  follow its evolution. The divergence in \eqref{2.2:bs} is treated
  using the cut-off method with a Gaussian vorticity profile
  \cite{saffman1992vortex}: to determine the local contribution to the velocity
  field at $\vec{X}_i^n$ from the neighboring segments $[\vec{X}_i^{n-1},
  \vec{X}_i^{n}]$ and $[\vec{X}_i^n, \vec{X}_i^{n+1}]$, we replace the two
  segments by an arc of circle passing through $[\vec{X}_i^{n-1}, \vec{X}_i^n,
  \vec{X}_i^{n+1}]$ and use the cut-off formula. The discretised expressions
  that we shall use for the induced velocity are given in
  \cite{venegas2019generalized}. When local contributions are not taken into
  account, the number of segments required to obtain a good approximation of the
  induced velocity increases by a factor $O(2\pi\rho/a)$, where $\rho$ is the
  local curvature radius \cite{venegas2019generalized}.

  Our objective is to find a vortex structure that is steady in the frame of the
  rotor. This condition of steadyness can be written as
  \begin{equation}\label{eq:steady}
    \left(\vec{U}(\vec{X}_j)  - \vec{U}^F  \right)\times \vec{T}_j = 0
  \end{equation}
  where $ \vec{U}^F$ is the frame velocity, which simply indicates that vortices
  are moving along the vortex structure. In practice, we shall write the
  steadiness condition as a system of ordinary equations. In a cylindrical
  frame, it can be written for each vortex as
  \begin{equation}\label{eq20}
    \frac{\dd r_j}{\dd \zeta} = U_r(\vec{X}_j),
    \quad\quad
    \frac{\dd \theta_j}{\dd \zeta} = \Omega(\vec{X}_j) - \Omega^F,
    \quad\quad
    \frac{\dd z_j}{\dd \zeta} = U_z(\vec{X}_j) - U_z^F,
  \end{equation}
  where $(r_j(\zeta),\theta_j(\zeta),z_j(\zeta))$ are the radial, angular, and
  axial coordinates of the $j$-th vortex,  $(U_r, r_j \Omega, U_z)$ the
  corresponding velocity components, while $\Omega^F$ and $U_z^F$ are the
  angular and axial velocity of the rotor frame. Each vortex curve is
  parametrized by $\zeta$. These systems are to be solved with boundary
  conditions on the rotor, at $\zeta=0$, where the position of each vortex is
  prescribed, and far-field boundary conditions at $\zeta \rightarrow \infty$.

  These last boundary conditions are not trivial as the far-field is a priori
  unknown. Therefore, our first task is to characterize the far-field.


  \section{Far-field solutions}
  \label{sec:3}

  We are interested in steady solutions created by the emission of two
  closely-spaced co-rotating vortices from a rotating blade tip under an
  external axial flow. Based on the observation of the tip vortices
  \cite{leishman2016principles}, one may naturally expect close to the rotor, a
  global contraction of the structure in a climbing helicopter regime, and a
  global expansion in a wind turbine regime. In the far-field, a quasi-uniform
  regime is expected. However, this regime is not as simple as for a single-tip
  vortex. We shall see that while not perfectly uniform, the vortices deform but
  exhibit a certain periodicity induced by their mutual interaction. The
  description of this far-field can be analyzed using the method introduced in
  \cite{venegas2019generalized} for counter-rotating vortices. These
  solutions are defined only by the geometrical parameters introduced in the
  following section.

  \subsection{Simplified approach}

  \subsubsection{Geometrical parameters}

  \begin{figure}[h]
    \hspace{3.5cm}(a) \hspace{0.30\linewidth} (b) \hfill   ~\\
    \includegraphics[scale=1.0]{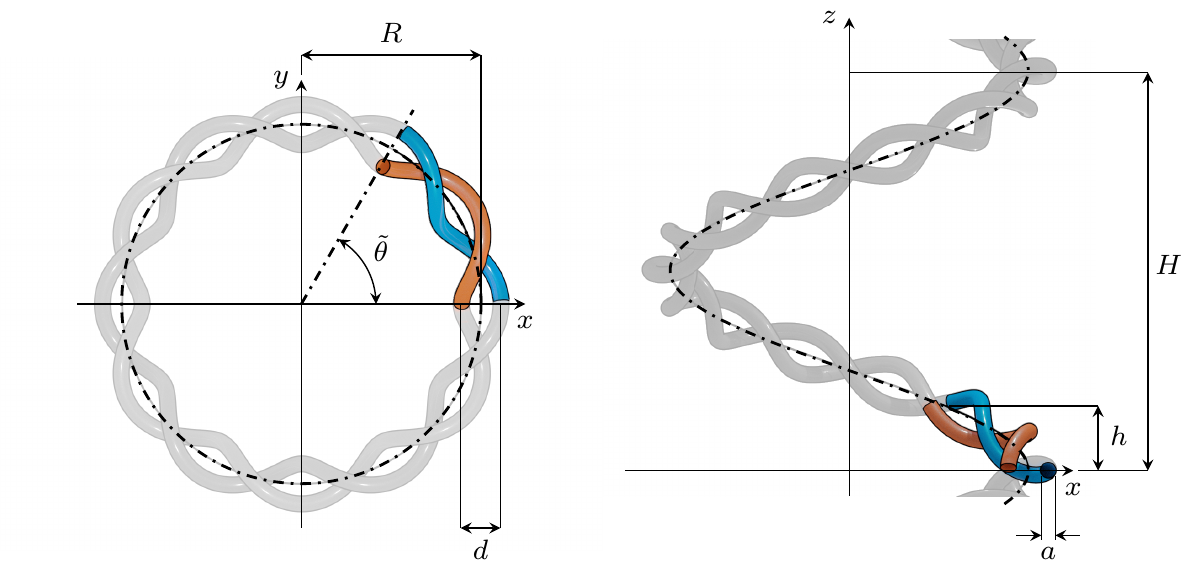}
    \caption{Geometric parameters of closely spaced tip vortices: separation
    distance $d$, radius $R$, vortex core size $a$, axial pitches $H$ and $h$,
    and twist parameter $\beta=\pm 2\pi/\tilde{\theta} = H/h$.}
    \label{fig1:intro}
  \end{figure}
  \begin{figure}[h]
    (a) \hspace{0.3\linewidth} (b) \hspace{0.3\linewidth} (c) \hfill~ \\
    \includegraphics[width=\linewidth]{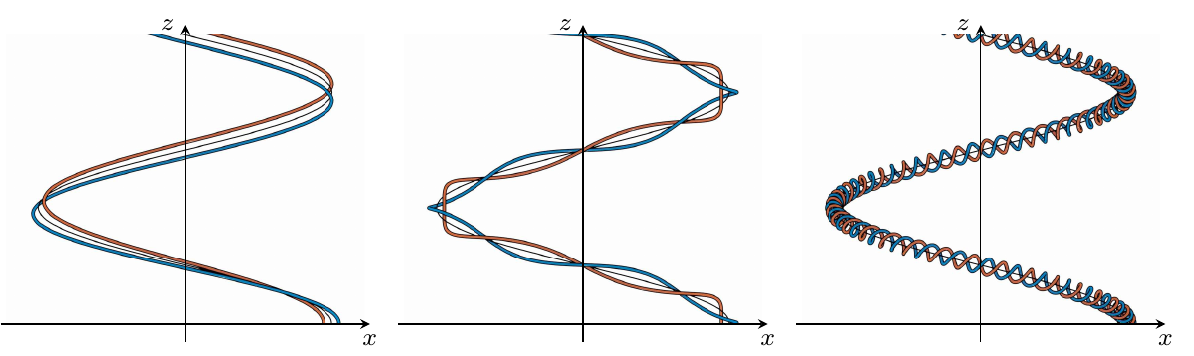}
    \caption{%
      Geometry described by \eqref{eq1.1} for $H=10$, $R=10$, $d=1$ and: (a)
      $h_\tau=128$ (or $\beta \approx 0.5$); (b) $h_\tau=32$ (or $\beta\approx
      2$); and (c) $h_\tau=2$ (or $\beta \approx 31.8$).
    }
    \label{fig2:intro}
  \end{figure}

  In the present situation, the far-field structure is expected to be close to a
  double-helix inscribed on a ``larger'' underlying helix. Such idealized
  structure is defined by the geometric parameters identified in figure
  \ref{fig1:intro}. The double-helix has a radius $d/2$ and pitch $h_\tau$
  (corresponding to an axial pitch $h$) while the large helix, noted
  $\mathscr{H}$, has a radius $R$ and axial pitch $H$. Another important
  parameter is the ratio $\beta \equiv H/h = 2\pi/ \tilde{\theta}$ (see figure
  \ref{fig1:intro}). Depending on the values of $H$, $R$, $h_\tau$, and $d$, the
  double-helix structure may describe:
  {\it (i)} a leapfrog-type pattern, where vortices
  trade places every $1/\beta$ turns  (figure \ref{fig2:intro}a);
  {\it (ii)} a relatively
  sparse braid (figure \ref{fig2:intro}b) ; or
  {\it (iii)} a dense `telephone cord'-type
  pattern (figure \ref{fig2:intro}c). We are typically in situation {\it (i)}
  when $\beta <1$ and  $h_\tau/d >10$, and in situation {\it (iii)} when
  $h_\tau/d < 5$  and $\beta>5$.

  The curve $\vec{X}_0$ described by $\mathscr{H}$ is defined in a Cartesian
  frame as a function of the angular coordinate $\theta_0$ by
  \begin{subequations}\label{eq1.0}
    \begin{eqnarray}
      x_0 &=& R \cos\theta_0
      \\
      y_0 &=& R\sin\theta_0
      \\
      z_0 &=& H\theta_0/(2\pi)
    \end{eqnarray}
  \end{subequations}
  while the curves $(\vec{X}_i$, $i=1,2)$ described by the double-helix are defined by
  \begin{subequations}\label{eq1.1}
    \begin{eqnarray}
      x_i &=& x_0 +
      (d/2)\cos\phi_i\cos\theta_0 - c_\tau (d/2)\sin\phi_i \sin\theta_0
      \\
      y_i &=& y_0 +
      (d/2)\cos\phi_i\sin\theta_0 + c_\tau (d/2)\sin\phi_i \cos\theta_0
      \\
      z_i &=& z_0 - c_\kappa(d/2)\sin\phi_i
    \end{eqnarray}
  \end{subequations}
  where $\phi_1 \equiv \beta\theta_0$ and $\phi_2 \equiv \beta\theta_0 + \pi $
  define the orientation of the helices relative to the chord plane.
  The torsion coefficient $c_\tau$ and the curvature coefficient $c_\kappa$ of
  $\mathscr{H}$ are given by,
  \begin{equation}
    c_\tau \equiv \frac{|H|}{\sqrt{4\pi^2R^2 + H^2}} ~ ,
    \quad \quad\quad\quad
    c_\kappa \equiv \frac{2\pi R}{\sqrt{4\pi^2R^2 + H^2}} ~.
  \end{equation}
  We also note that $h_\tau$ and $h$ are related with each other by the relation
  \begin{equation}
    h_\tau = \frac{h}{c_\tau}.
  \end{equation}

  The above structure satisfies the following form of spatial periodicity: it is
  invariant by the double operation  $z\rightarrow z+h$ and
  $\theta_0\rightarrow\theta_0+\tilde{\theta}$. Moreover, since we consider
  vortices of equal core size $a$ and circulation $\Gamma$, both vortices are
  deemed as interchangeable, such that our periodic domain is further reduced to
  a domain of axial length $h/2$ and azimuthal angle $\tilde{\theta}/2$. In the
  following, we keep this property to obtain the steady solutions. In particular, we
  assume that there is a single location over an axial period $h/2$ where both
  vortices are at the same azimuth. We chose this azimuth to define the radius
  $R$ and separation distance $d$. The axial period $h$ then defines a mean
  axial pitch while $H$ is obtained from $H= 2\pi h / \tilde{\theta}$. In
  addition to the core size $a$ that is assumed constant, these length scales can be used to define four
  dimensionless parameters
  \begin{equation}\label{eq3.1:params}
    R^* \equiv \frac{R}{d}, \quad\quad
    H^* \equiv \frac{H}{d}, \quad\quad
    h_\tau^* \equiv \frac{h_\tau}{d}, \quad\quad
    \varepsilon \equiv \frac{a}{d} ,
  \end{equation}
  that will characterize our solutions in the far-field.

  \subsubsection{Characteristics of the moving frame }
  \label{sec:frame-approx}

  A vortex ring and a helical vortex are examples of vortex structures that move
  in space at a constant speed without deformation \cite{lamb1945hydrodynamics}.
  The double-helix structure is also bound to move by the induction of velocity
  it generates. For this structure, one can naturally assume that the
  self-induced velocity is predominantly composed of:
  \begin{itemize}
    \item a translation of velocity $u_t$ and rotation of angular velocity
    $\omega$ around the double-helix axis, i.e., $\mathscr{H}$.
    \item a translation of velocity $U_z$ and rotation of angular velocity
    $\Omega$ around the large helix axis, i.e., the vertical axis.
  \end{itemize}

\begin{figure}
    \hspace{0.0cm}(a) \hspace{0.475\linewidth} (b) \hfill ~\\
    \includegraphics[scale=1, trim=0 0 0 0, clip]{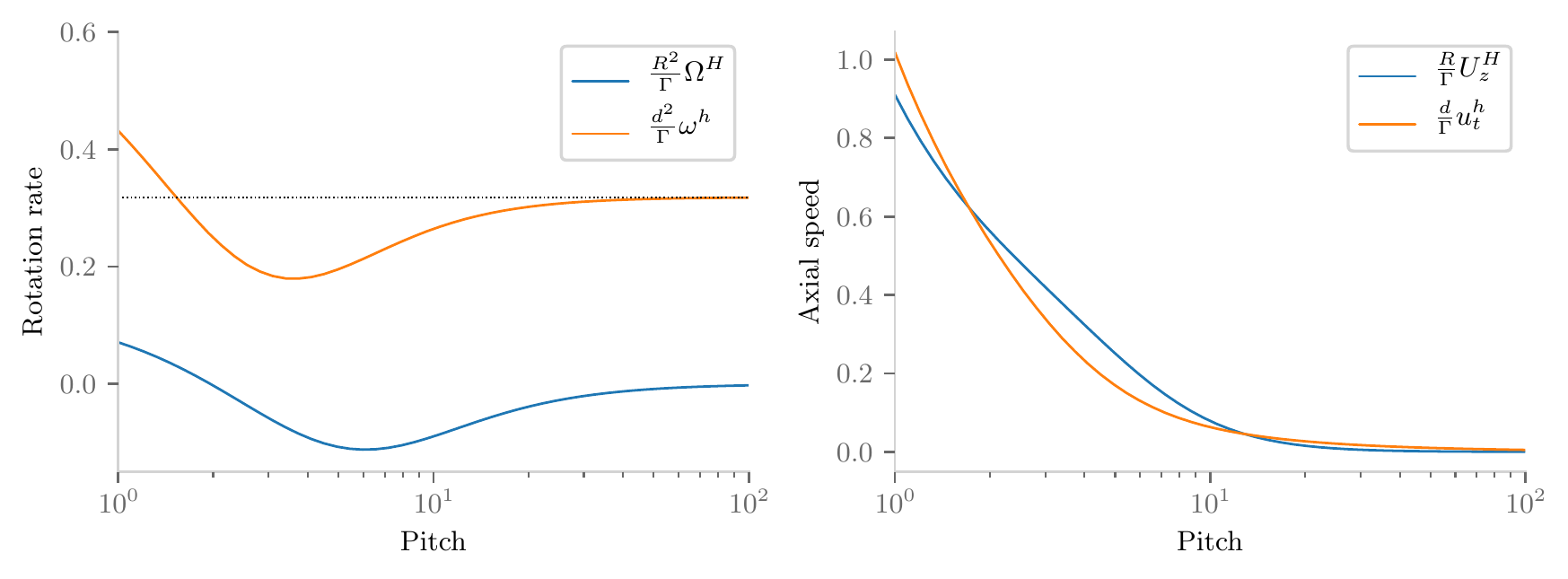}
    \caption{
      Dimensionless rotation rate and axial speed of the vortex elements on
      a helix of radius $R$, circulation $2\Gamma$ and core size $d=0.1 R$
      as a function of the pitch $H/R$ (in blue), and a
      double-helix of radius $d/2$, circulation $\Gamma$ and core size $a=0.1
      d$ as a function of the pitch $2h_\tau/d$ (in orange).
      The dotted line in (a) indicates the value $d^2\omega^h/\Gamma = 1/\pi$
      for $h_\tau\gg1$.
    }
    \label{fig3:approx}
  \end{figure}

  Under this assumption, we can show there exists a unique frame rotating
  and translating along the vertical axis where the double-helix geometry shown
  in figure \ref{fig1:intro} remains steady. The idea is to use the property
  that a helix of pitch $H$ is unperturbed by a rotation of angular velocity
  $\Omega^a$  and translation of axial velocity $U_z^a$ provided that
  \begin{equation}
    U_z^a/\Omega^a = \pm H/(2\pi),
  \end{equation}
  where the sign is $+$ for right-handed helices, and $-$ for
  left-handed helices.  The self-rotation of the double-helix can then be
  cancelled by adding a motion of velocity $u_t^a = \mp h_{\tau} \omega /(2\pi)
  $ along the double-helix axis. This motion corresponds to an axial rotation of
  angular velocity $c_\kappa u_t^a/R$ plus an axial translation of velocity
  $c_\tau u_t^a$. These velocities add up to the self-induced velocities of the
  large helix. Both sums can be cancelled if one chooses the frame velocities
  such that
  \begin{equation}\label{eq1.6}
    \Omega^F = \Omega + \frac{c_\kappa u_t}{R}   \mp \frac{h}{H}\omega ~,
    \quad\quad\quad\quad
    U_z^F = U_z + c_\tau u_t \mp \frac{c_\tau h_\tau}{2\pi} \omega ~.
  \end{equation}
  These expressions will be used as initial guess values for the frame velocity
  in the numerical procedure described in the next section. If the vortices have
  a circulation $\Gamma$, we shall use
  \begin{itemize}
    \item for the self-induced motion ($\omega$ and $u_t$) of the double-helix,
    the values obtained by the cut-off method with a Gaussian vortex core of
    radius $a$ for a straight double-helix of circulation $\Gamma$ and the
    same geometrical parameters (pitch $h_{\tau}$, radius $d/2$).
    \item for the self-induced motion ($\Omega$ and $U_z$) of the large helix
    $\mathscr{H}$, the values obtained by the cut-off method with a Gaussian
    vortex core of ``effective'' radius $d/2$ for a helix of circulation
    $2\Gamma$, pitch $H$ and radius $R$.
  \end{itemize}
  Typical values for these velocities are provided as function of the
  non-dimensional pitch in figure \ref{fig3:approx}. In the following, we shall
  refer to \eqref{eq1.6} with these estimates as the rectilinear approximation
  for the frame velocity.

  \subsection{Numerical solutions}

  In the previous section, a prescribed shape for the solutions has been
  assumed. This hypothesis will turn out to be too strong. We shall see that the
  helices must deform to be compatible with the self-induced motion. For the
  numerical procedure, we use the separation distance $d$ and the vortex
  circulation $\Gamma$ to normalize time and length scales. The solutions will
  then only depend on the four dimensionless parameters $R^*$, $H^*$, $h_\tau^*$
  and $\varepsilon$ introduced in \eqref{eq3.1:params}. To reduce the size of
  the parameter space, only $H^*$ and $h_\tau^*$ will be varied. In most cases,
  we shall use $R^*=9.5$ and $\varepsilon= 0.1$,
  for which the filament approach is expected to provide reliable results.

  \subsubsection{Governing equations in helical coordinates}

  \begin{figure}
    \hspace{2.25cm}(a) \hspace{0.4\linewidth} (b) \hspace{0.38\linewidth} ~\\

    \centering
    \includegraphics[width=0.9\columnwidth, trim=0 0 0 0, clip]{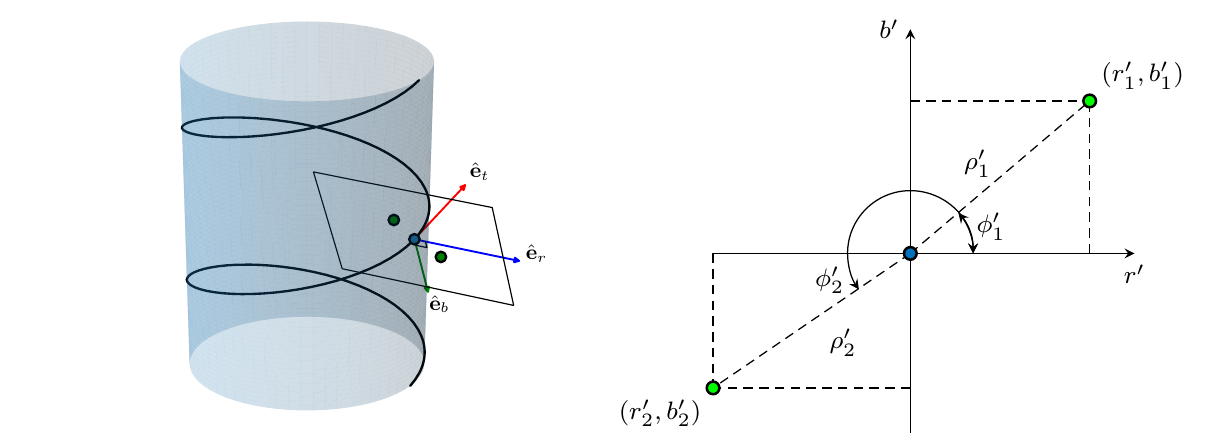}
    \caption{
    (a) Local helical basis following $\mathscr{H}$ and (b) 2-D coordinates in
    the $(r', b')$ plane.
    }
    \label{fig4:frame}
  \end{figure}

  To describe the far-field, it is useful to introduce a local Frenet-Serret
  basis $(\bhat{e}_{r0}, \bhat{e}_t, \bhat{e}_b)$ based on $\mathscr{H}$ and
  centered at $\vec{X}_0$ as illustrated in figure \ref{fig4:frame}a. The
  position vector $\vec{X}_j$ will then be defined by its local coordinates
  $r'_j = (r_j-R^*)$ and $b'_j$ in the local perpendicular basis $(\bhat{e}_{r0},
  \bhat{e}_b)$, see figure \ref{fig4:frame}b.
  Developing $\vec{T}_j$ in terms of $(\bhat{e}_{r0}, \bhat{e}_t, \bhat{e}_b)$, the
  steadiness condition \eqref{eq:steady} may be written as
  \begin{eqnarray}\label{eq:14}
    \frac{dr'_j}{d\theta_0} = \left( \frac{R^*}{c_\kappa} + c_\kappa r'_j \right)
    \frac{U_{r_0} - U_{r_0}^F}{U_t - U_{t}^F } + c_\tau b'_j
    \quad\quad
    \frac{db'_j}{d\theta_0} = \left( \frac{R^*}{c_\kappa} + c_\kappa r'_j \right)
    \frac{U_b - U_{b}^F}{U_t - U_{t}^F } - c_\tau r'_j.
  \end{eqnarray}

  Because of the spatial periodicity and interchangeability conditions, the
  calculation domain is comprised between $\theta_0=0$ and
  $\theta_0=\tilde{\theta}/2$, with
  \begin{equation}
  \tilde{\theta} = 2\pi/\beta
  \end{equation}
  being the half-period of the vortex pair. Our description is completed by
  the following boundary conditions
  \begin{equation}
    \begin{cases}
      r'_1(0) = r'_2(\tilde{\theta}/2) = +\frac{1}{2},
      \\
      r'_2(0) = r'_1(\tilde{\theta}/2) = -\frac{1}{2},
    \end{cases}
    \quad\quad
    \begin{cases}
      b'_1(0) = b'_2(\tilde{\theta}/2) = 0,
      \\
      b'_2(0) = b'_1(\tilde{\theta}/2) = 0 .
    \end{cases}
  \end{equation}

  Equation \eqref{eq:14} is solved numerically as a non-linear minimization
  problem. Starting from an initial guess for $(r'_j, b'_j)$ obtained from
  \eqref{eq1.1}
  \begin{equation}\label{eq3.6}
    r'_1 = \frac{1}{2}\cos(\beta\theta_0),
    \quad\quad
    b'_1 = \frac{1}{2}\sin(\beta\theta_0),
    \quad\quad
    r'_2 = \frac{1}{2}\cos(\beta\theta_0 + \pi),
    \quad\quad
    b'_2 = \frac{1}{2}\sin(\beta\theta_0 + \pi),
  \end{equation}
  and an initial guess for $\Omega^F$ and $U_z^F$ obtained from \eqref{eq1.6},
  one applies an iterative procedure. At step $n$ of this procedure, one
  integrates \eqref{eq:14} between $\theta_0=0$ and $\theta_0=\tilde{\theta}/2$
  to compute the new values of $\Omega^F$ and $U_z^F$ which satisfy the boundary
  conditions. Then, the next iteration of $(r'_j, b'_j)$ is computed from the
  residual value of \eqref{eq:14} using Newton's method. Steps are repeated
  until the current and subsequent iterations converge. To fix the spatial
  resolution, we check for grid convergence using the primitive variables, local
  curvature coefficient and frame velocities. In practice, for most
  calculations we use $p_n = 48$ segments per period.

  \subsubsection{Geometry of the solutions}

  \begin{figure}
    (a) \hspace{0.45\linewidth} (b) \hfill ~\\
    \includegraphics[width=\columnwidth, trim=15 0 0 80, clip]{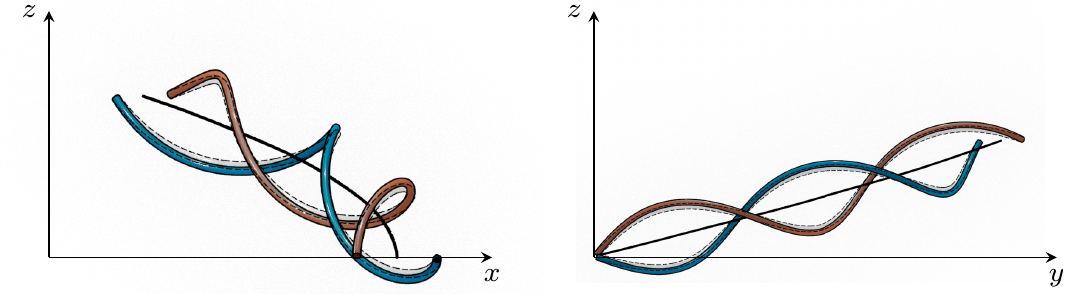}
    \caption{
      Close-up of the deformed vortex structure for $H^*=15$ and $h_\tau^*=5$
      over a single period. Numerical solution (in colour) superimposed to the
     undeformed  initial  guess (in dashed lines).
    }
    \label{fig5:deformation}
  \end{figure}

\begin{figure}
    \hspace{1.0cm}(a) \hspace{0.4\linewidth} (b) \hfill ~\\
    \includegraphics[width=\columnwidth, trim=0 0 0 0, clip]{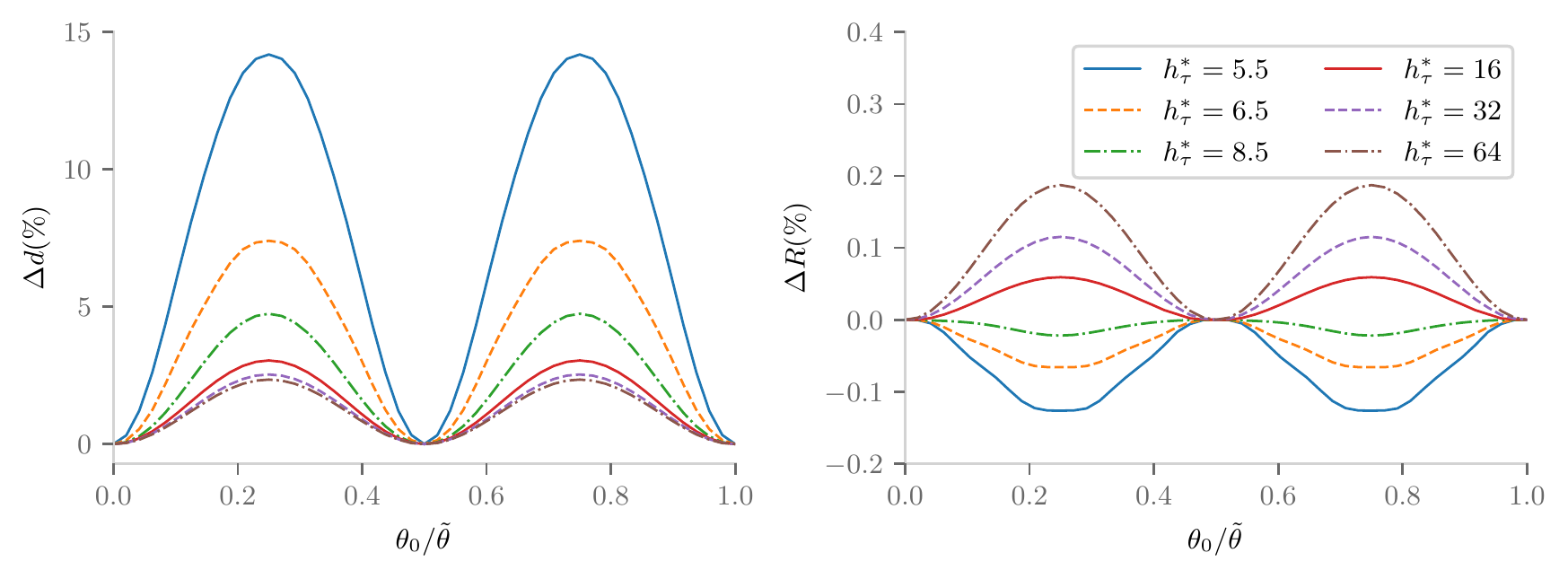}

    \hspace{1.0cm}(c) \hspace{0.4\linewidth} (d) \hfill ~\\
    \includegraphics[width=\columnwidth, trim=0 0 0 0, clip]{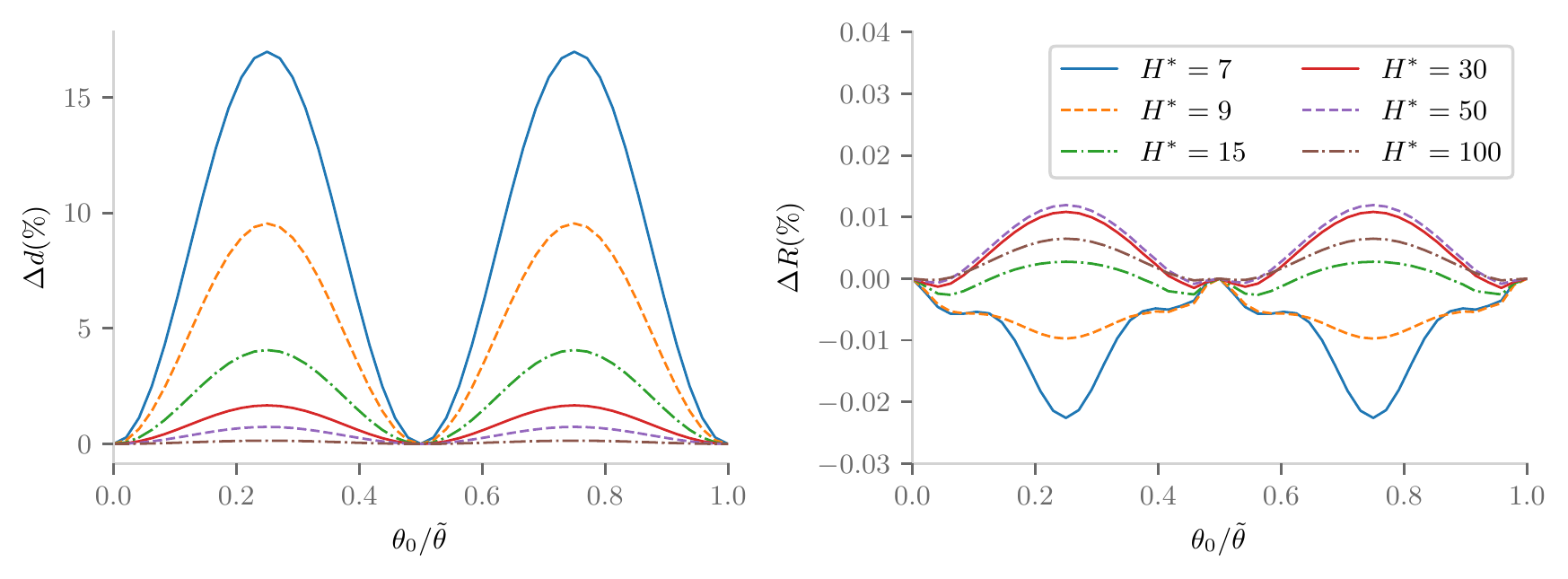}
    \caption{
      (a,c) Change in the separation distance $\Delta d$ and (b,d) mean radius
      $\Delta R$ as function of $\theta_0$ for: (a,b) $H^*=15$ and different
      $h_\tau^*$; and (c,d) $h_\tau^*=10$ and different $H^*$.
    }
    \label{fig6:deformation}
  \end{figure}

\begin{figure}
    \hspace{1.0cm}(a) \hspace{0.4\linewidth} (b) \hfill ~\\
    \includegraphics[width=\columnwidth, trim=0 0 0 10, clip]{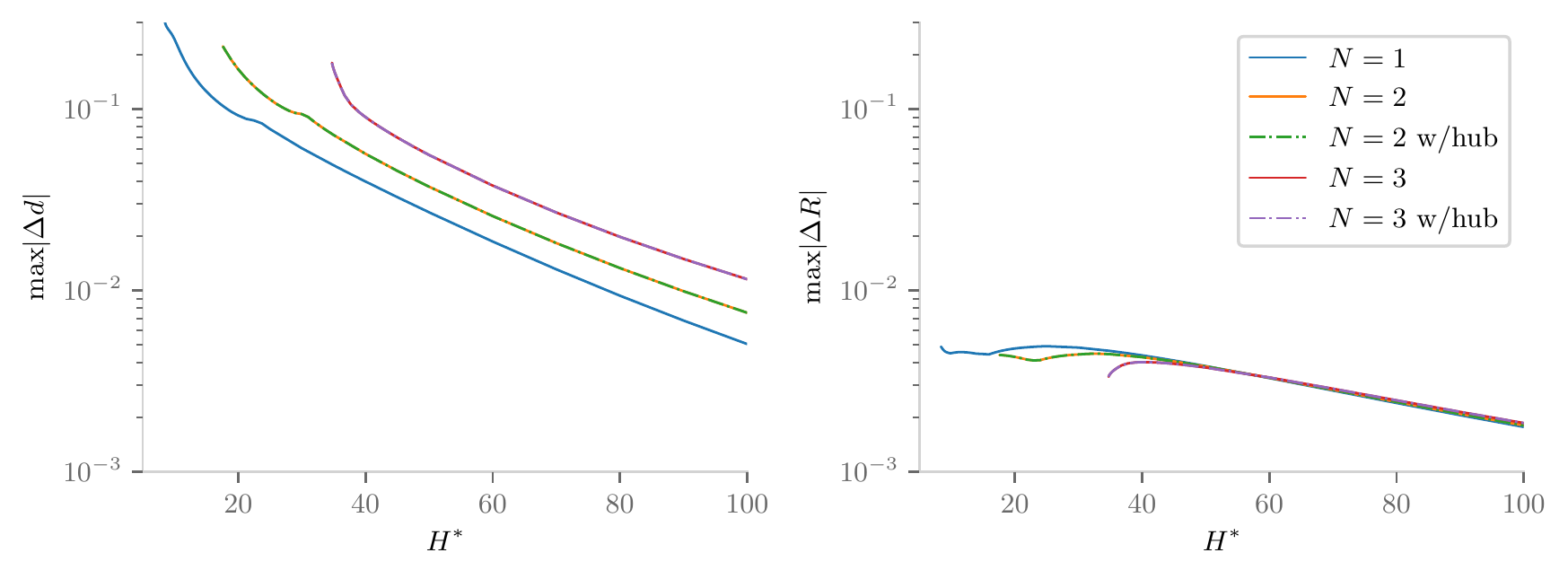}
    \caption{
      Maximum change in (a) the separation distance $\vert \Delta d \vert$ and (b) mean
      radius $\vert\Delta R \vert$ as function of $H^*$ for $h_\tau^*=2$ and $N$ vortex
      pairs with and without a straight hub vortex of circulation $-2N\Gamma$.
    }
    \label{fig7:deformation}
  \end{figure}

  In this section, we characterize the geometry of the solutions, especially
  their departure from the (undeformed) initial guess. In general, solutions
  keep the overall form of the approximated solutions, but the helical
  symmetries are now broken (figure \ref{fig5:deformation}).  Both the
  double-helix structure and the large underlying helix deform.  Vortex segments
  placed in the first and second half-periods are usually drawn in opposite
  directions. This combined motion affects the local curvature, as it
  straightens vortex filaments around the nodes, and bends them around the
  anti-nodes. Some useful indicators of this deformation are the relative
  changes of the separation distance and of the mean radius defined by
  \begin{equation}\label{eq:def:deltaxi}
    \Delta d(\theta_0) = \sqrt{(r'_1 -r'_2)^2 + (b'_1 -b'_2)^2} -1 , \quad\quad
    \Delta R(\theta_0) = \frac{\sqrt{x_1^2 + y_1^2} + \sqrt{x_2^2 + y_2^2}}{2R^*} - 1,
  \end{equation}
  respectively.
  A positive (resp. negative) $\Delta d$  indicates that the distance between
  the two vortices is increasing (resp. decreasing). Similarly, a positive
  (resp. negative) $\Delta R$ indicates an increase (resp. decrease) of the mean
  radius. By construction, $\Delta d(0)=\Delta d(\tilde{\theta}/2)= \Delta
  R(0)=\Delta R(\tilde{\theta}/2)= 0$. As seen in figure
  \ref{fig6:deformation}a-b, both quantities are shown to vary over a single
  period and their evolution varies with the type of solution. For leapfrogging
  wakes {\it (i)}, $\Delta d$ remains relatively constant while $\Delta R$ is
  positive with local maxima near $\tilde{\theta}/4$. For sparsely braided wakes
  {\it (ii)}, $\Delta d$ reaches a plateau while $\Delta R$ becomes close to
  zero. For densely coiled wakes {\it (iii)}, $\Delta R$ may take positive and
  negative values, while the maximum in $\tilde{\theta}/4$ becomes a local
  extremum. The maximum $|\Delta d|$ and $|\Delta R|$ initially contracts,
  before increasing once again as $H^*$ decreases (figure
  \ref{fig7:deformation}). Below this point, the mutually-induced velocity,
  which drives the rotation of the vortex pair, is no longer the dominant
  contribution to the total induced velocity and numerical convergence becomes
  increasingly hard to obtain for this family of solutions.

\begin{figure}
    \hspace{1.5cm} (a) \hfill (b) \hfill (c) \hfill\hspace{1.5cm} ~\\

    \centering
    \includegraphics[width=0.8\columnwidth, trim=0 0 0 0, clip]{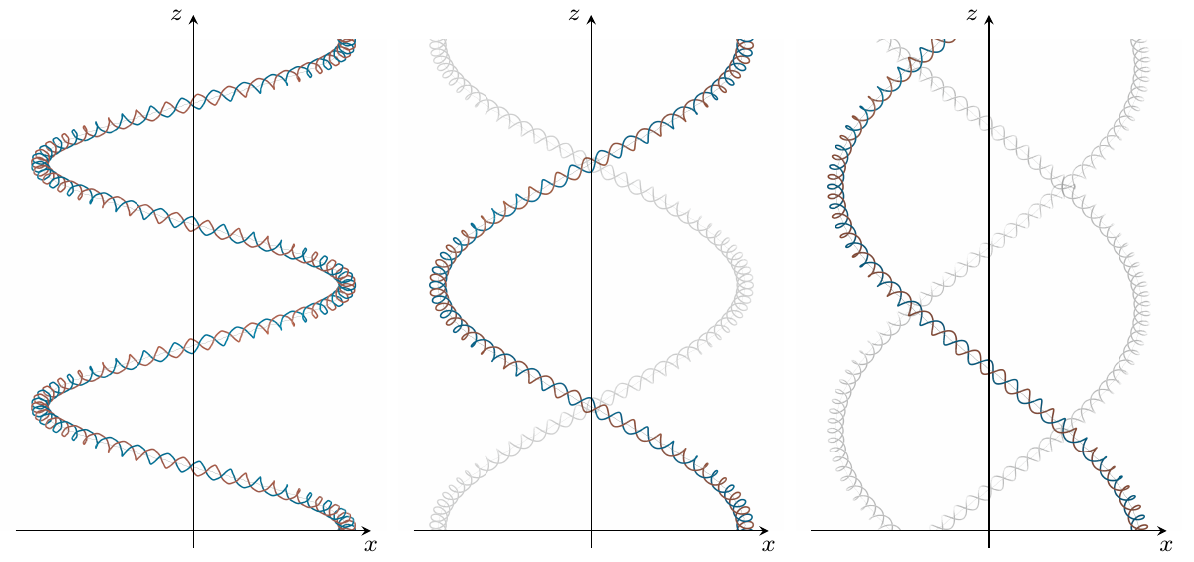}
    \caption{
    Deformed structures for (a) $N=1$, (b) $N=2$ and (b) $N=3$ vortex pairs
    for $H^*/N=15$ and $h_\tau^*=2$.
    }
    \label{fig8:deformation}
  \end{figure}

  For applications, it is interesting to compare these solutions to a similar
  configuration composed of $N$ pairs of vortices of circulation $\Gamma$. In
  this context, this wake geometry would represent the far-field in a Joukowski
  rotor wake model \cite{zhukovskiui1929theorie}, but produced by a $N$-bladed
  tip-splitting rotor. Figure \ref{fig8:deformation}a (resp.
  \ref{fig8:deformation}b) displays the wake geometry obtained for two (resp.
  three) vortex pairs for $h_\tau^*=2$. In general, for equal $(H^*, h_\tau^*)$,
  deviations from \eqref{eq1.1} are found to increase with $N$ (figure
  \ref{fig7:deformation}). However, if we base our comparison on the separating
  distance between neighbouring spires, the deformation actually decreases with
  $N$. A similar effect has been reported for counter-rotating helical vortices
  \cite{venegas2020}. Finally, we note that including a straight vortex hub of
  circulation $-2N\Gamma$ was shown to have little effect on the wake geometry.
  Unless otherwise stated, in the sequel, we shall focus on the case for $N=1$
  without a hub vortex.

  \subsubsection{Frame and tangential velocities}

  \begin{figure}
    \hspace{1.0cm}(a) \hspace{0.4\linewidth} (b) \hfill ~\\
    \includegraphics[width=\columnwidth, trim=0 0 0 5, clip]{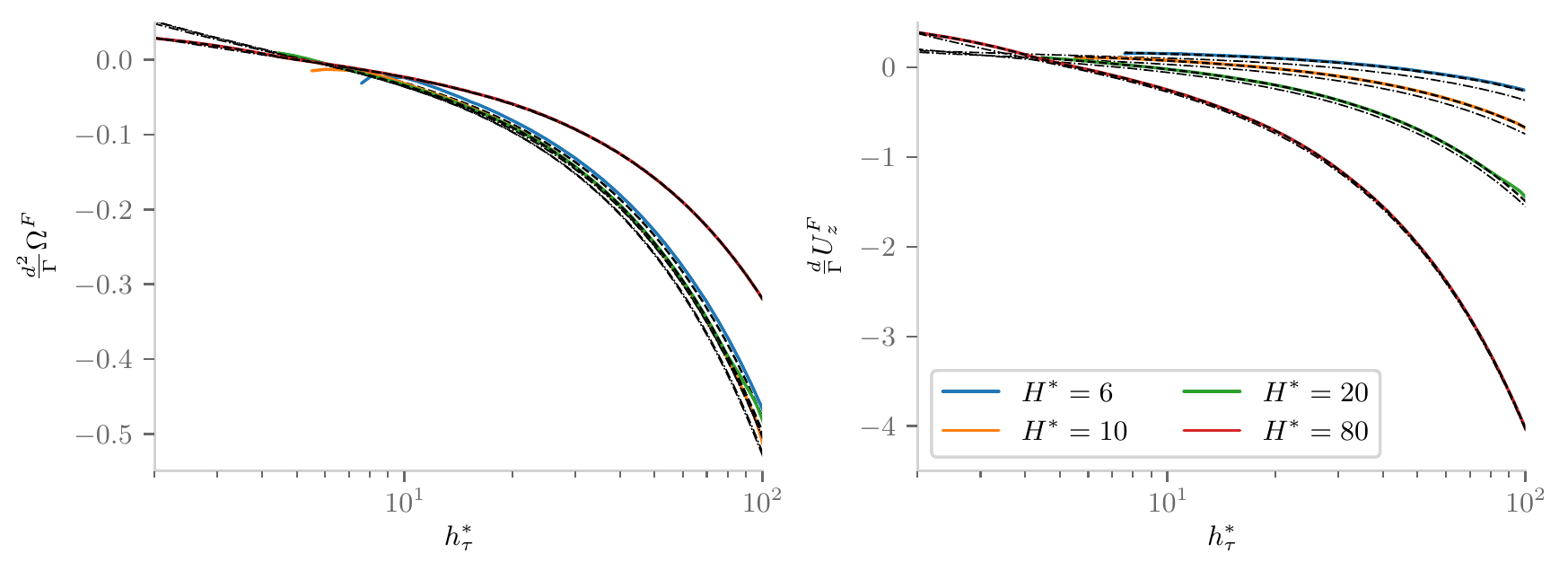}
    \caption{
      Frame velocities as function of $h_\tau^*$. Numerical solutions (in solid
      lines) are compared to the rectilinear approximation (in dash-dotted
      lines) and to a first order approximation based on the undeformed initial
      guess (in dashed lines).
    }
    \label{fig9:frames}
  \end{figure}
  \begin{figure}
    \hspace{1.0cm}(a) \hspace{0.4\linewidth} (b) \hfill ~\\
    \includegraphics[width=\columnwidth, trim=0 0 0 10, clip]{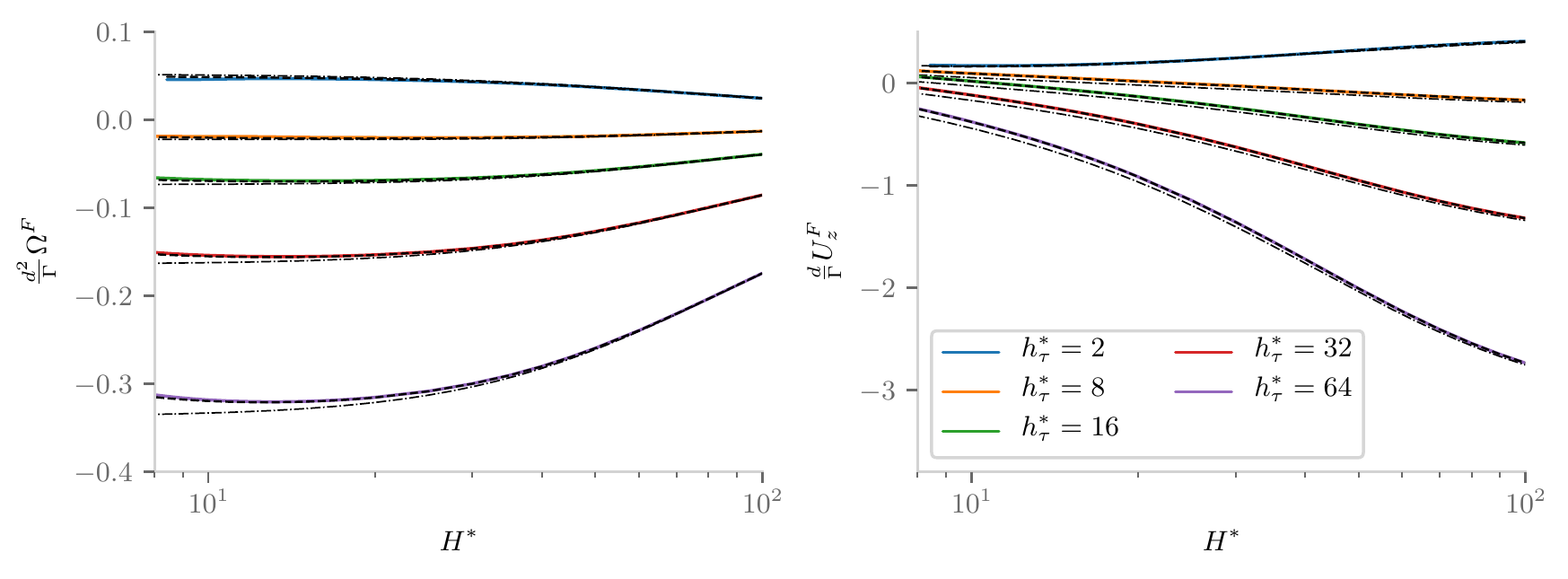}
    \caption{
      Frame velocities as function of $H^*$. Numerical solutions (in solid
      lines) are compared to the rectilinear approximation (in dash-dotted
      lines) and to a first order approximation based on the
      undeformed initial guess (in dashed lines).
    }
    \label{fig10:frames}
  \end{figure}
  \begin{figure}
    \hspace{1.0cm}(a) \hspace{0.45\linewidth} (b) \hfill ~\\
    \includegraphics[width=\columnwidth, trim=0 0 0 0, clip]{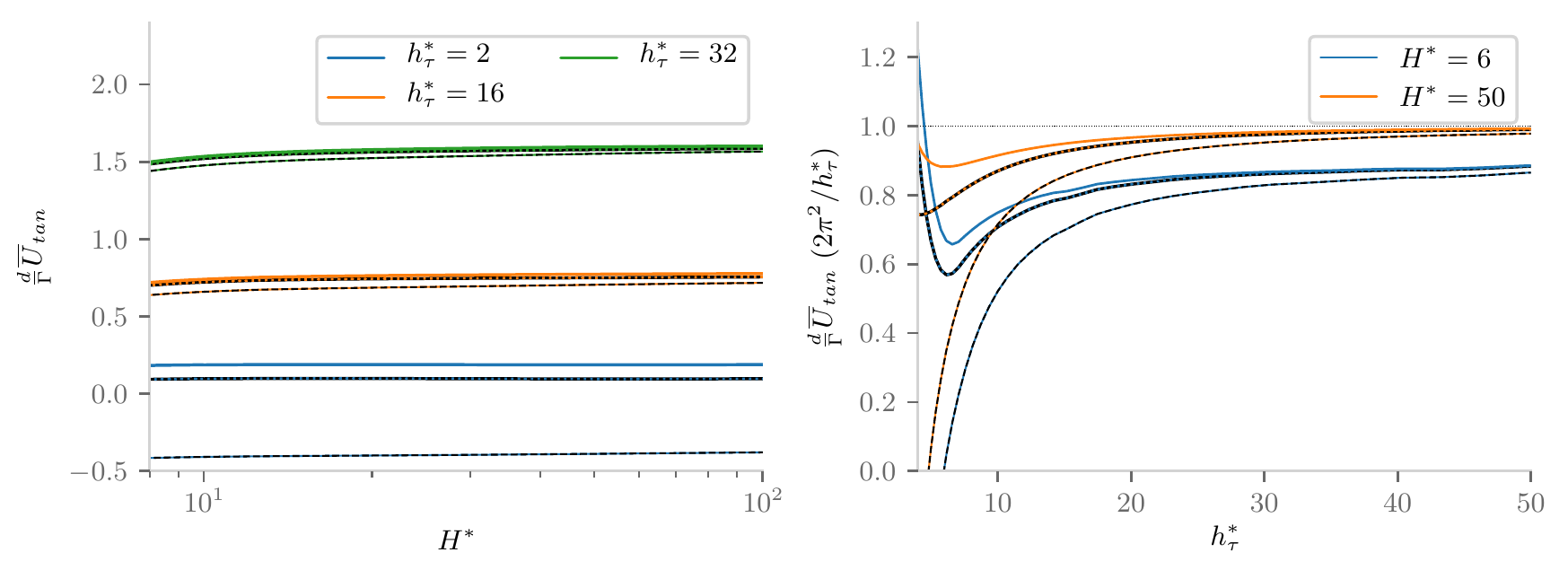}
    \caption{
      Evolution of $ \overline{U}_{tan} $ (in solid lines) and $
      \overline{U}_{t} $ (in dotted lines) as function of (a) $H^*$ and (b)
      $h_\tau^*$. Numerical solutions are also compared to the velocity
      associated with the moving frame $ \overline{U}^F_{t} $ (in
      dashed lines).
    }
    \label{fig11:frames}
  \end{figure}

  The numerical procedure also provides the frame velocities $\Omega^F$ and
  $U_z^F$ where the structure is steady.  These quantities are plotted as a
  function of $h_{\tau}^*$ (resp. $H^*$) in figure \ref{fig9:frames} (resp.
  figure \ref{fig10:frames}). They are compared to the rectilinear approximation
  obtained in \S\ref{sec:frame-approx} (in dash-dotted lines) and to a first
  order approximation obtained by numerically computing the frame velocities
  with an undeformed solution (in dashed lines). Both approximations work well
  for large $H^*$ and large $h_\tau^*$, i.e., when self-induction is expected to
  be small. The first order approximation, which takes into account the mutual-
  and self-induced velocities but neglects the deformation, is naturally better
  and provides accurate estimates whenever the deformations are weak.

  Now let us consider the velocities relative to the moving frame. By
  construction, the vortex elements are advected along the stationary vortex
  structure, where positive tangential velocity $U_{tan}$ indicates an advection
  in the positive axial direction and {\it vice versa}. In general, values of
  $U_{tan}$ are positive (resp. negative) for $h^*_\tau<0$ (resp. $h^*_\tau>0$).
  The tangential velocity varies with respect to $\theta_0$ identically for the
  internal and external vortices but is shifted by a half-period. Variations are
  generally small with respect to mean velocity $\overline{U}_{tan}$. As seen on figures
  \ref{fig11:frames}a and \ref{fig11:frames}b, the mean velocity displays a
  linear dependency on $h_\tau$ with only a weak dependency on $H^*$ for
  $H^*<10$. Figure \ref{fig11:frames}a also shows that, whatever $h_{\tau}^*$ and
  $H^*$, the main contribution to the tangential velocity is the translation
  speed $U_t$ along $\mathscr{H}$. We suspect that this comes from the large
  value of $R^*$ that we have considered  ($R^*=9.5$).

  It is also interesting to compare $U_{t}$ to the velocity associated
  with the moving frame
  \begin{equation}
    U^F_{t} = c_\tau U_z^F + c_\kappa R^* \Omega^F
  \end{equation}
  since the difference between the two is issued from vortex induction. For
  leapfrogging and weakly braided configurations ($h_\tau^* > 10$), most of the tangential
  velocity is associated with the moving frame and the role of vortex induction
  is generally small. Note in particular that when $H^*$ is also large, we
  recover the estimate $\overline{U}_{tan} \approx  h_\tau \Gamma / (2 \pi^2
  d^2) $ obtained from the rectilinear approximation in this limit. By contrast,
  for densely coiled wakes, such as $h_\tau^*=2$, $\overline{U}_{tan}$ and
  $\overline{U}_{t}$ can be significantly larger than the frame velocity and
  even point opposite directions (figure \ref{fig11:frames}b).

  \subsubsection{Properties of the induced flow}

  \begin{figure}
    \quad(a) \hfill (b) \hfill (c) \hfill \quad~ \\
    \includegraphics[width=\columnwidth, trim=0 0 0 0, clip]{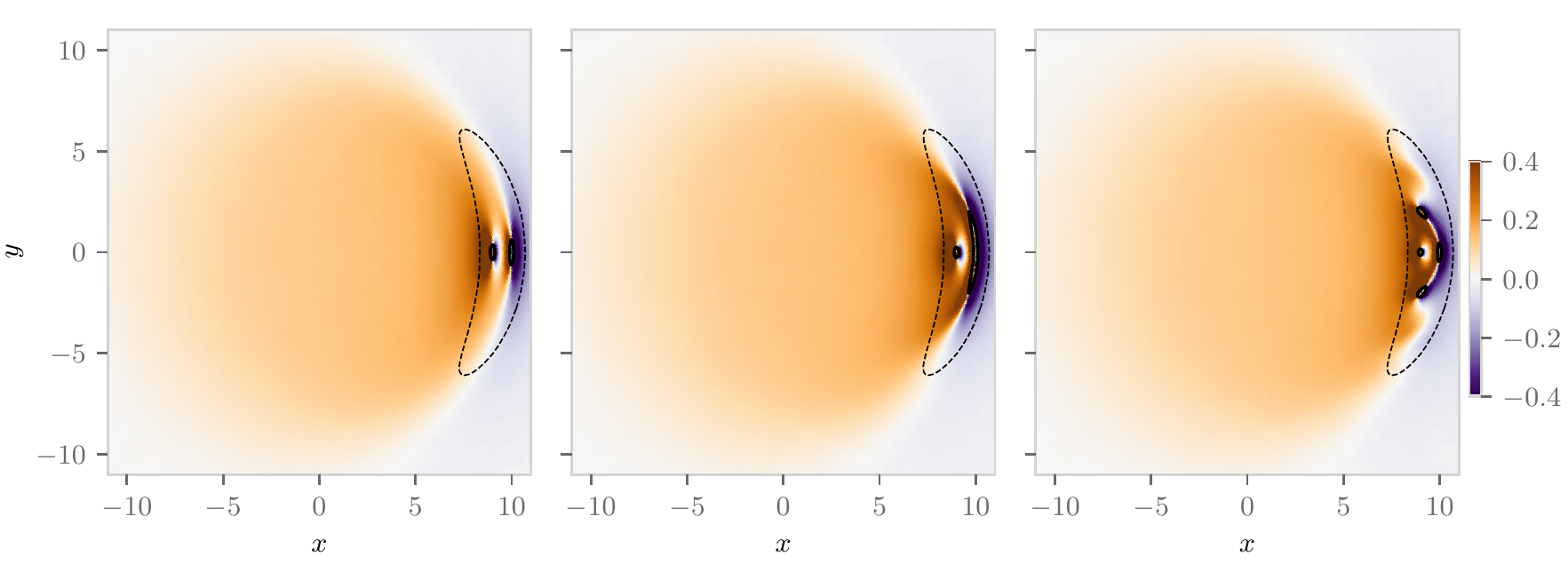}

    \includegraphics[width=\columnwidth, trim=0 5 0 0, clip]{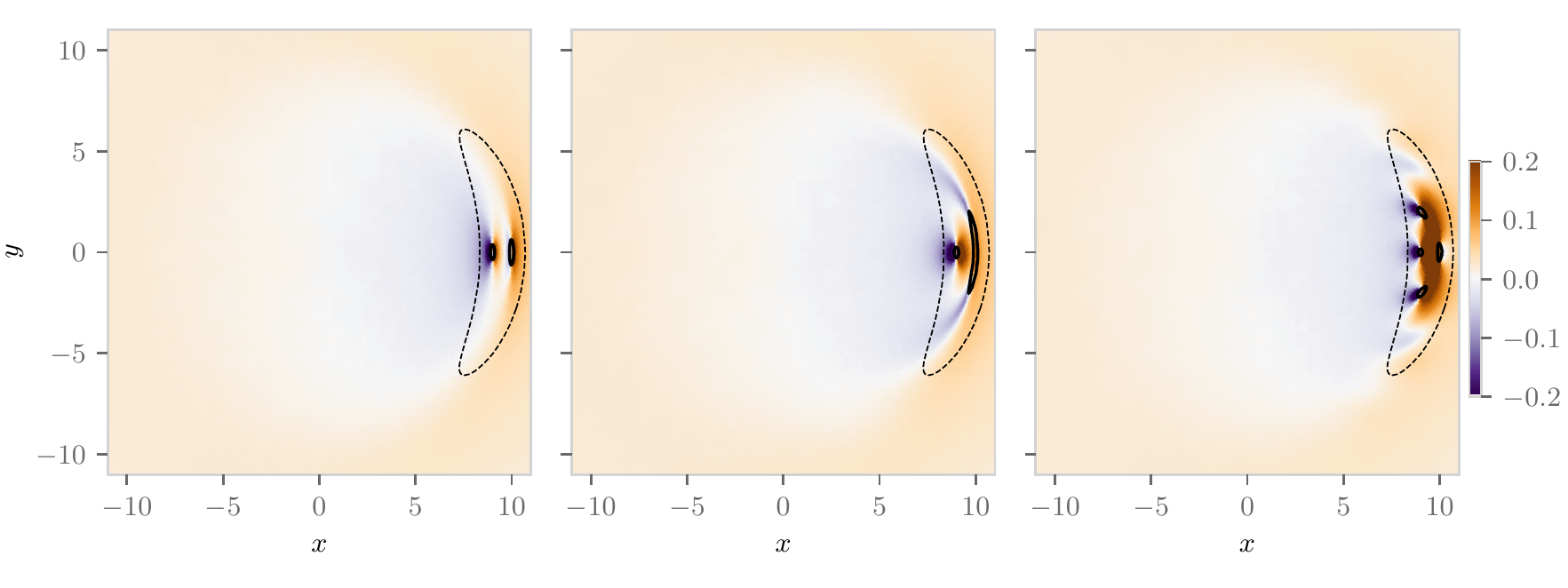}
    \caption{
      Axial velocity $\frac{d}{\Gamma} U_z $ (top) and angular velocity
      $\frac{d}{\Gamma} U_\theta $ at the plane $(x,y,z=0)$ for $H^*=15$. (a)
      Leapfrogging wake for $\beta=1$, (b) sparsely braided wake for $\beta=4$,
      and (c) densely coiled wake for $\beta=16$. Solid lines indicate the
      intersection with the vortex cores, while dashed lines correspond to a
      helical vortex tube enclosing the twin-vortex.
    }
    \label{fig12:fields}
  \end{figure}
  \begin{figure}
    \hspace{1.0cm}(a) \hspace{0.45\linewidth} (b) \hfill ~\\
    \includegraphics[width=\columnwidth, trim=0 5 0 0, clip]{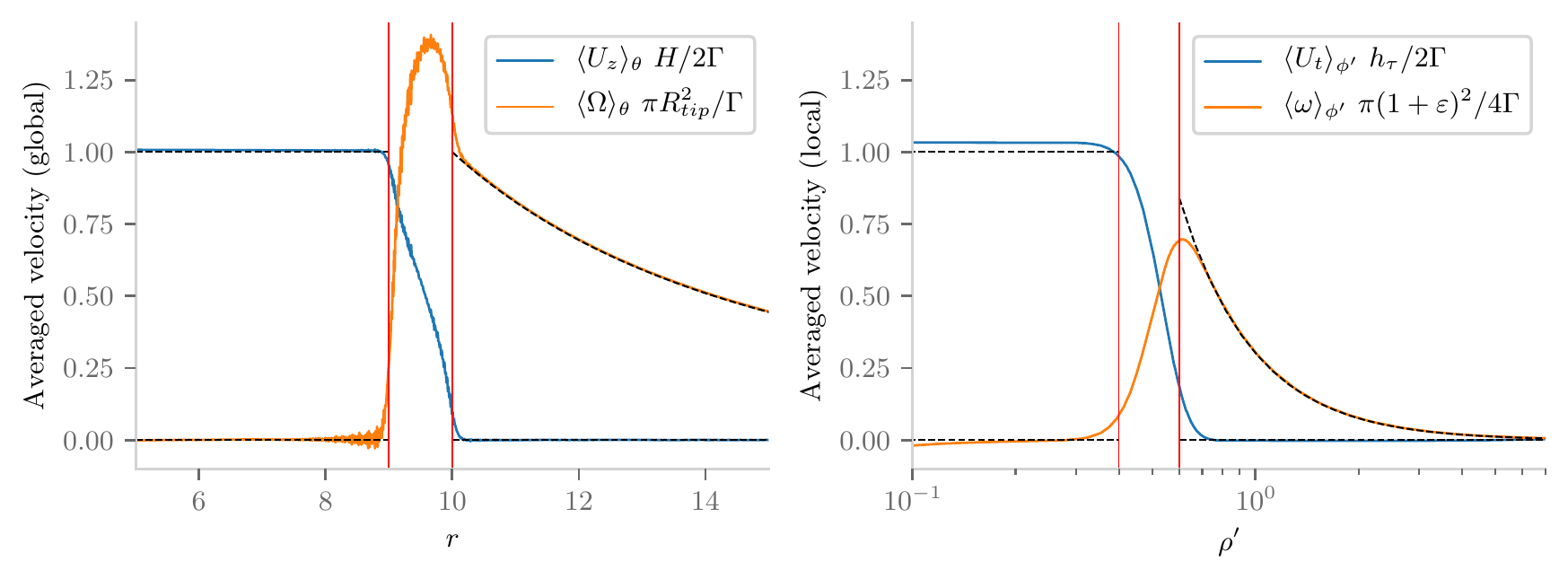}
    \caption{
      Azimuthally averaged velocity profiles for $H^*=15$ and $h_\tau^* =2$.
      Fig. (a) (resp. (b)) Rotation rate and axial velocity component in the
      global cylindrical (resp. local helical) frame. Dashed lines correspond to
      expressions \eqref{eq:HardinH} (resp. \eqref{eq:Hardinh}).
    }
    \label{fig13:fields}
  \end{figure}

  For applications, it is useful to know the velocity field induced by the
  vortex structure. In this section, the induced flow is provided in cross
  planes and compared to the predictions obtained from the Kawada-Hardin
  solutions for perfect helices \cite{kawada1936induced, hardin1982velocity,
  fukumoto2015contribution}. Figures \ref{fig12:fields}a-c show the axial and angular velocity
  components taken at the plane $z=0$ (where both vortices have the same
  azimuth) for a leapfrogging, sparsely braided and densely coiled structures,
  respectively. In these plots, we subtract the frame velocities such that the
  velocity field vanishes far from the center. Vortex cores intersect the plane
  on numerous occasions resulting in inhomogeneous velocity and vorticity
  distributions. For densely coiled structures, this results in regions of very
  intense angular velocity and opposite signed vorticity. Nevertheless, away
  from the cores (outside the region enclosed in dashed lines)
  the induced velocity is reminiscent to that of a helical vortex.

  To highlight this similarity, consider the azimuthally averaged velocity
  profiles presented in figure \ref{fig13:fields}a. Inside the region $R^* - \frac{1}{2}
  < r < R^* + \frac{1}{2}$, the angular velocity displays a maximum around $r=R^*$, while
  the axial velocity can be roughly approximated by a linear function. Outside
  the same region the azimuthally averaged profiles
  approach the ideal profiles of a uniform helix obtained from the Kawada-Hardin
  solutions \cite{kawada1936induced, hardin1982velocity,
  fukumoto2015contribution}
  \begin{equation}
    \left\langle U_z^{H} \right\rangle_\theta =
    \begin{cases}
      2\Gamma/H & \mbox{ if } r<R^* -1/2
      \\
      0 & \mbox{ if } r>R^* +1/2
    \end{cases} ~;
    \quad\quad
    \left\langle \Omega^{H} \right\rangle_\theta =
    \begin{cases}
      0 & \mbox{ if } r<R^* -1/2
      \\
      \Gamma/(\pi r^2) & \mbox{ if } r>R^* +1/2
    \end{cases}
    .
    \label{eq:HardinH}
  \end{equation}

  Close to the vortex cores, the velocity field can be better understood by
  projecting the velocity components into the local helical frame $(\rho',
  \phi')$. This is shown in figure \ref{fig13:fields}b,
  where the $\phi'$-averaged profiles also approach the mean profiles of a
  perfect double-helix locally aligned with $\mathscr{H}$
  \begin{equation}
    \left\langle U_t^{h} \right\rangle_{\phi'} =
    \begin{cases}
     2\Gamma/h_\tau & \mbox{ if } \rho'< (1 - \varepsilon)/2
      \\
      0 & \mbox{ if } \rho'> (1 + \varepsilon)/2
    \end{cases}~;
    \quad\quad
    \left\langle \omega^{h} \right\rangle_{\phi'} =
    \begin{cases}
      0 & \mbox{ if } \rho'< (1 - \varepsilon)/2
      \\
      \Gamma/(\pi {\rho'}^2)  & \mbox{ if } \rho'> (1 + \varepsilon)/2
    \end{cases}.
    \label{eq:Hardinh}
  \end{equation}

  The above results highlight the dual nature exhibited by this vortex
  structure. Solutions for closely spaced co-rotating helical vortices issue
  from a balance between contributions at different scales with long-distance
  effects primarily governed by the large-scale helix and more local features
  defined by the double-helix geometry.

  \section{Near-field solutions}
  \label{sec4}

  In the previous section, we have described the solutions in the far-field. In
  this section, we consider the vortex structure close to the rotor plane. The
  objective is not to perform a complete parametric study but to illustrate how
  these solutions can describe the wake of a tip-splitting rotor under different
  flight regimes.

  \subsection{Numerical method}

  There is one notable difference with respect to the analysis performed in the
  previous section. Instead of fixing the geometric parameters to find the
  associated frame velocities, now we solve the inverse problem. That is, we fix
  the operating conditions, i.e., the rotation rate $\Omega^F = \Omega_R$ and
  external velocity $U_z^F = -U_{\infty}$, and compute the corresponding wake
  geometry. Now the prescribed geometrical parameters are the radial coordinates
  $r=R_{fin}=R_0-d_0/2$ and $r=R_{tip}=R_0+d_0/2$ where the vortices are
  emitted on the rotor blade (located at $\theta_0=0$ and $z=0$), and the vortex
  core size $a$. As previously, we define four non-dimensional parameters
  \begin{equation}
    \lambda \equiv \frac{R_{0} \Omega_R}{U_{\infty}},
    \quad\quad
    \eta \equiv \frac{\Gamma}{R_{0}^2 \Omega_R},
    \quad\quad
    R_0^* \equiv \frac{R_{0}}{d_{0}},
    \quad\quad
    \varepsilon_0^* \equiv \frac{a}{d_0} ,
  \end{equation}
  where $\lambda$ is known as the tip-speed ratio and $\eta$ represents the
  relative vortex strength. As in \cite{venegas2020}, we choose a convention
  such that $\eta$ remains positive, while $\lambda$ may change sign. Wind
  turbine regimes will always correspond to positive values of $\lambda$. By
  contrast, helicopters may correspond to either negative values of $\lambda$ in
  climbing flight or positive values of $\lambda$ in descending flight. In the
  present study, only the parameters  $\eta$ and $\lambda$ are varied. In most
  cases, $\varepsilon_0^{*}$ and $R_0^{*}$ are fixed to 0.1 and 9.5,
  respectively. Note that these parameters are different from the parameters
  $R^*$ and $\varepsilon^*$ defined above for the far-field, as the mean radius
  and the vortex separation distance in the far-field are now the results of the
  calculation.

  \begin{figure}
    \includegraphics[width=0.6\linewidth]{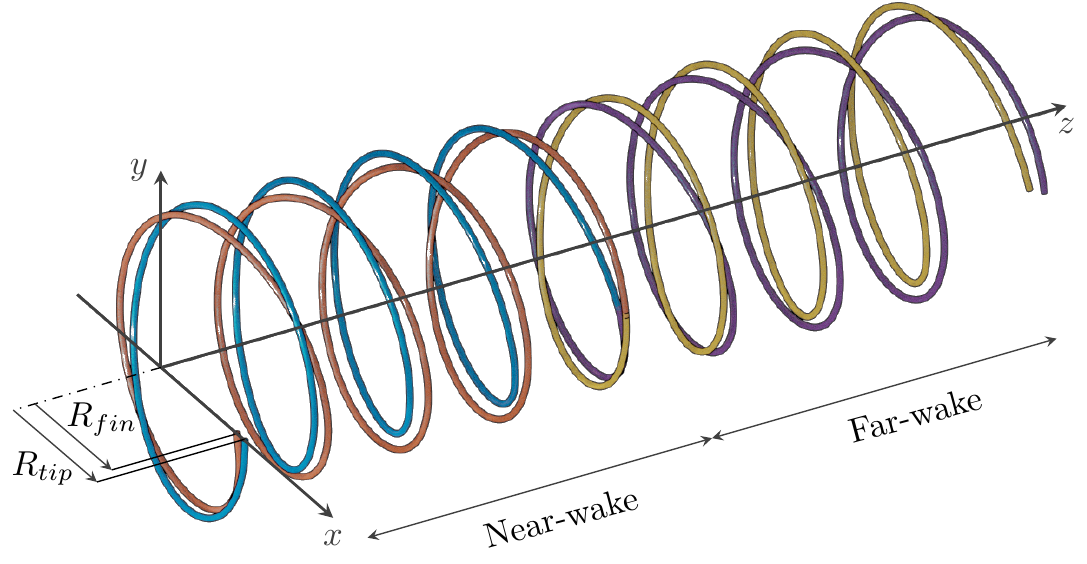}
    \caption{
      Schematic representation of the semi-infinite case: two co-rotating
      vortices are emitted from a rotor blade at radial positions $R_{fin}$ and
      $R_{tip}$, respectively. The calculation domain and the prescribed
      far-wake are shown in different colors.
    }
    \label{fig13:scheme}
  \end{figure}

  The near-field solution satisfies (\ref{eq20}) with the boundary condition at
  $\zeta=0$: $r_1=R_0^*-\frac{1}{2}$ and $r_2=R_0^*+\frac{1}{2}$ and $\theta_j=z_j=0$. As we go
  away from the rotor, it should match a far-field solution. To implement this
  condition, we follow the numerical procedure used by \citet{venegas2020} for
  Joukowski's rotor wake model. We decompose the induced velocity into
  contributions from the near-field and far-field. Contributions from the
  far-field are modeled by imposing that after a certain distance from the rotor
  plane, the wake adopts the geometry of a far-field solution (see figure
  \ref{fig13:scheme}). The geometrical parameters of this far-wake solution are
  estimated from the near-field solution at the end of calculation domain. The
  computational domain must be large enough for the wake to develop and match
  the far-field. In practice, the size of the domain and of the corresponding
  far-wake range from $30\pi$ to $60\pi$.

  \subsection{ Description of the steady solutions }
  \label{sec4.2}

  \begin{figure}
    (a) \hfill~ (b) \hfill~ \\
    \includegraphics[width=\columnwidth, trim=0 0 0 0, clip]{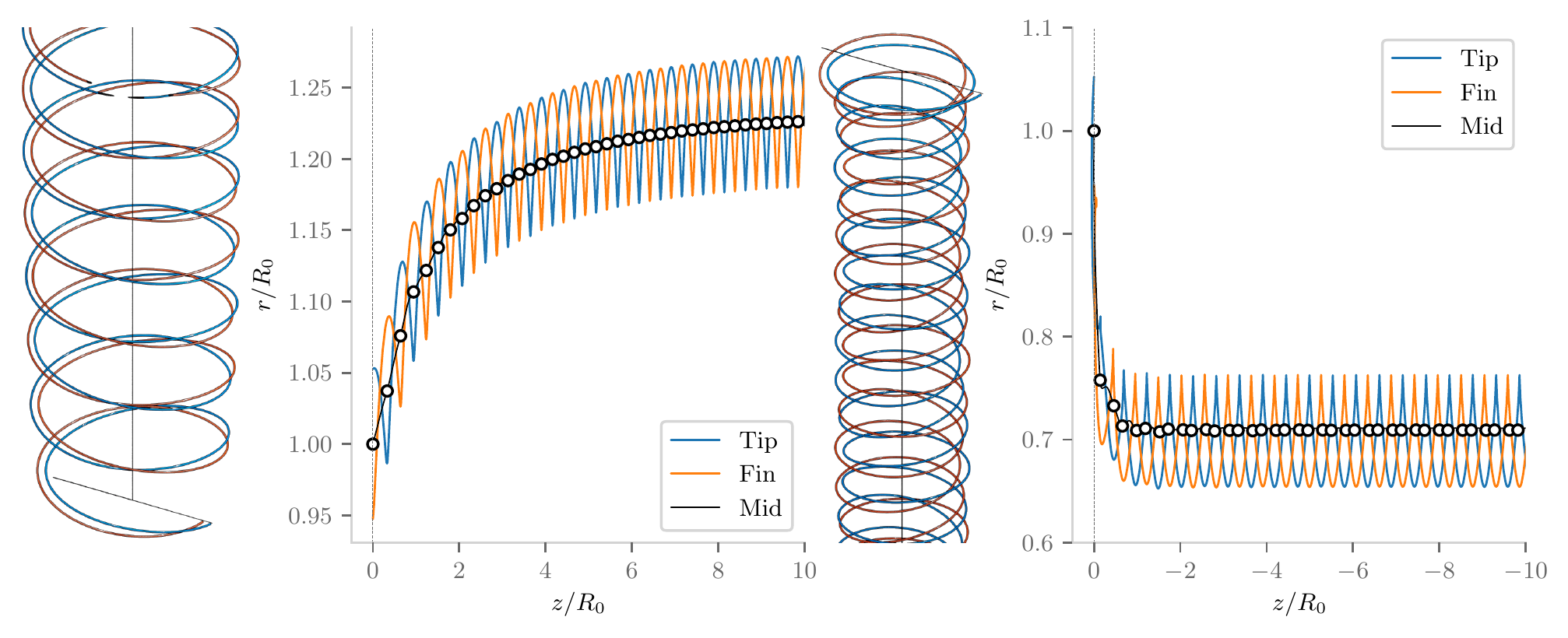}

    \caption{
    Schematic representation of the vortex wake and evolution of the radial
    coordinates $r_j$ as function of $z$ for $\eta=0.04$, 
     $R_0^{*}=9.5$, $\varepsilon_0^*=0.1$.
    (a) $\lambda=5.5$ in a wind turbine regime; (b)
    $\lambda=40$ in a helicopter descending flight regime. Markers indicate the
    positions where both vortices have the same azimuth.
    }
    \label{fig14:rotor}
  \end{figure}

  Figure \ref{fig14:rotor} displays the wake structure with the corresponding
  radial coordinates as function of the axial coordinate $z$. Figure
  \ref{fig14:rotor}a (resp. \ref{fig14:rotor}b) is representative of the wind
  turbine  (resp. helicopter) regime, which displays a radial
  expansion (resp. contraction), while the two vortices trade places
  continuously. Simultaneously, the separation distance may also contract or
  expand due to the combined effects of the vortex pair and the rotor's
  presence. Additionally, the structure may exhibit local variations of both
  axial pitches. For clarity, we characterize the wake in terms of the
  geometric parameters introduced in \S\ref{sec:3}.

  \subsubsection{Evolution of the wake in the near-field }

  \begin{figure}
    \quad (a) \hfill (b) \hfill ~ \\
    \includegraphics[width=\columnwidth, trim=0 0 0 0, clip]{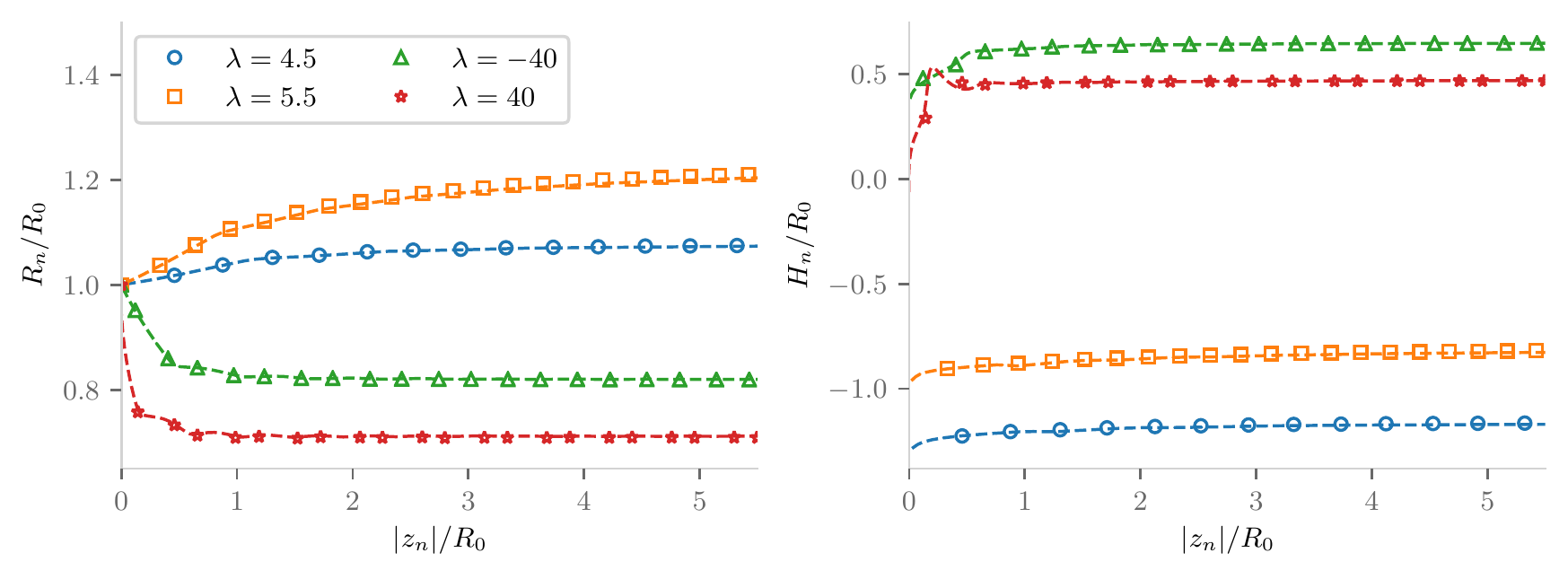}

    \caption{
    Evolution of (a) $R_n/R_0$, and (b) $H_n/R_0$ as function of $\vert
    z_n\vert/R_0 $ for different values of $\lambda$ for $\eta = 0.04$,
    $R_0^*=9.5$, $\varepsilon_0^{*}=0.1$. For each case, we display the
    corresponding evolution obtained for an `equivalent' helical vortex in
    dashed lines.
    }
    \label{fig16:nearwake}
  \end{figure}

  As in \S\ref{sec:3}, we introduce the curve $ \vec{X}_{mid}(\zeta)$
  equidistant to both vortices and the separation distance $d(\zeta)$ defined by
  \begin{equation}
    \vec{X}_{mid}(\zeta)=(\vec{X}_{tip}(\zeta)+\vec{X}_{fin}(\zeta))/2,
  \end{equation}
   \begin{equation}
    d(\zeta)=\lVert\vec{X}_{tip}(\zeta) - \vec{X}_{fin}(\zeta)\rVert.
  \end{equation}
  We also perform a series of discrete measurements at the points along the
  curve $(\zeta_n, n=1,2,\cdots)$ where the internal and external vortices have
  the same azimuth (indicated by solid marks in figure \ref{fig14:rotor}). In
  this way, the large helix $\mathscr{H}$ is characterized by a set of radii and
  pitches
  \begin{equation}\label{def:params:big}
    R_n \equiv r_{mid}(\zeta_n),
    \quad\quad\quad\quad\quad\quad\quad
    H_n \equiv \frac{z_{mid}(\zeta_{n+1}) - z_{mid}(\zeta_{n})}{\theta_{mid}(\zeta_{n+1}) - \theta_{mid}(\zeta_{n})},
    \quad
    \mbox{ for } n =  0,1,2,\cdots
  \end{equation}
  which converge towards the values at the far-field $R_\infty$ and $H_\infty$,
  respectively.

  Figure \ref{fig16:nearwake}a displays the evolution of $R_n/R_0$ as a function
  of $z(\zeta_n)/R_0$ for cases representative of wind turbine and helicopter
  regimes. Far-field solutions suggest that $\mathscr{H}$ can be modeled by a
  single tip vortex of circulation $2\Gamma$ and core size $d$ emitted from the
  radial coordinate $R_0$ under the same operating conditions. A good agreement
  is observed between the evolution of $R_n/R_0$ and the equivalent helix
  approximation. A similar agreement is observed between $H_n/R_0$ and the
  `equivalent' pitch in figure \ref{fig16:nearwake}b.

  Conversely, the rotation of the vortex pair is characterized by a set of
  mean separation distances and axial pitches
  \begin{equation}\label{def:params:small}
    d_n \equiv d (\zeta_n),
    \quad
    h_n \equiv \frac{z_{mid}(\zeta_{n+1}) - z_{mid}(\zeta_{n})}{\pi}
    \quad,
    \mbox{ for } n = 0,1,2,\cdots
  \end{equation}
  which converge towards the values at the far-field $d_\infty$ and $h_\infty$,
  respectively.
  Combining both sets of measurements, we obtain a set for the mean pitch and
  twist parameter
  \begin{equation}
    h_{\tau,n} \equiv h_n \frac{\sqrt{H_n^2 + 4\pi^2 R_n^2}}{H_n},
    \quad
    \beta_n \equiv \frac{H_n}{h_n}
    \quad,
    \mbox{ for } n = 1,2,\cdots
  \end{equation}
  which converge towards the values at the far-field $h_{\tau,\infty}$ and
  $\beta_\infty$.

  \begin{figure}
    \quad (a) \hfill (b) \hfill (c) \hfill ~ \\
    \includegraphics[width=\columnwidth, trim=0 0 0 0, clip]{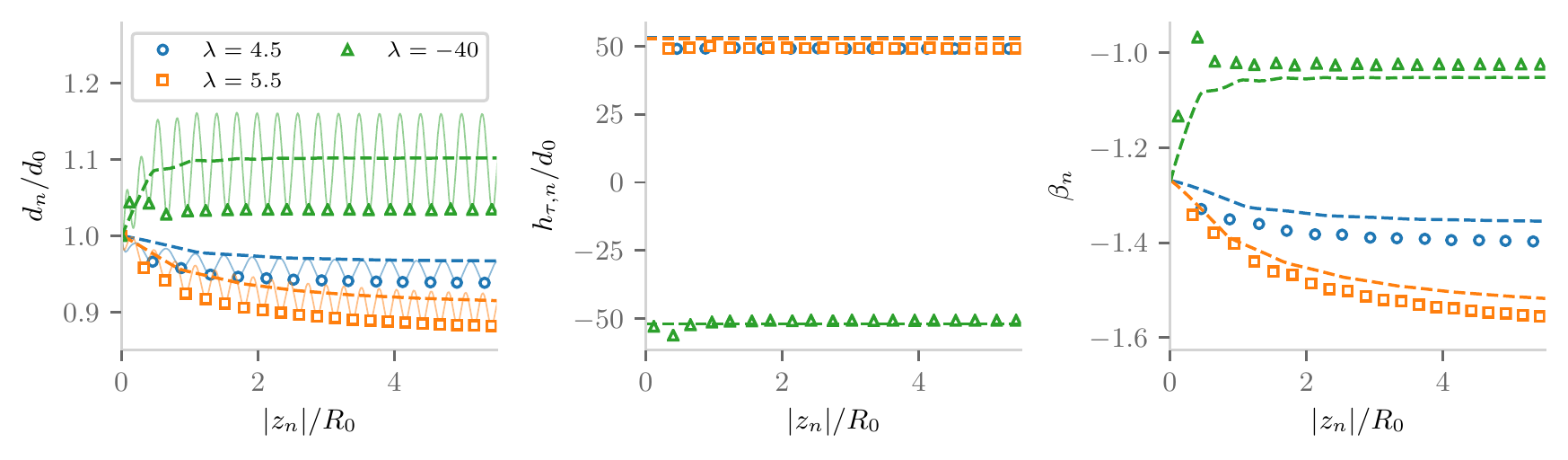}
    \caption{
      Evolution of (a) $d_n/d_0$, (b) $h_{\tau,n}/d_0$ 	and $\beta_n$ as a function of $z_n/R_0
      $ for the cases displayed in figure \ref{fig16:nearwake}.  Dashed lines
      in figures (a) and (b), correspond to \eqref{eq29:model} and
      \eqref{eq30:model}, respectively. For reference, the evolution of $d(z)/d_0$
      is shown in solid lines in fig. (a).
    }
    \label{fig15B:rotor}
  \end{figure}

  The evolution of these parameters is shown in figures \ref{fig15B:rotor}a-c.
  The first remarkable feature is a dynamical similarity between local
  parameters' evolution and those of the large helix. Variations are limited
  to the very near wake, and are rapid whenever those of the large helix are
  rapid. However, variations seem to be opposite for the small and the large
  helix: the separation distance $d_n$ increases (resp. decreases) when the
  large helix radius $R_n$ decreases (resp. increases). The local pitch
  $h_{\tau,n}$ is also found to have an opposite sign to $H_n$, meaning that the
  small and large helices are oriented differently. Additionally, both pitches
  remain constant in the wake. It is also important to note the large absolute
  value of $h_{\tau ,n}/d_0$, indicating that configurations are weakly twisted
  where self-induced effects are generally small. This has some
  consequences that can be used to derive a model explaining the variations of
  the local properties.

  The evolution of the separation distance $d_n$ can be understood by the
  following arguments. As the radius $R_n$ of the large helix varies, we expect
  variations of the advection speed $U_t$ of the vortices along $\mathscr{H}$.
  These variations are responsible of a local strain that tends to contract the
  vortex structure (that is, decrease $d$) when $U_t$ increases, and expand it
  when $U_t$ decreases. The variations of $d$ are then obtained by the
  conservation of the mass flux in the vortex stream tube which prescribes that
  \begin{equation}\label{eq:Utd}
    U_t(\zeta_n) d^2(\zeta_n)= cst.
  \end{equation}

  An estimate for $U_t(\zeta_n)$ can be obtained when the self-induction is
  weak. We have seen in \S\ref{sec:3} that in that case, $U_t(\zeta_n)$ is
  mainly provided by the frame velocity $U_t^F(\zeta_n) = \pm(U_\infty^2 + R_n^2
  \Omega_R^2)^{1/2}$ where the $\pm$ sign is defined by the sign of
  $U_{\infty}$. In terms of the non-dimensionalized parameters $\lambda$ and
  $\eta$, this gives
  \begin{equation}\label{eq28:model}
    U_{t}(\zeta_n)\frac{R_0}{\Gamma} \approx  \frac{\sqrt{1 + \lambda^2 R_n^2/R_0^2}}{\eta \lambda}
  \end{equation}
  Combining this expression with  (\ref{eq:Utd}), we obtain
  the following relation between $d_n$ and
  $R_n$
  \begin{equation}\label{eq29:model}
    \frac{d_n}{d_0} = \left[\frac{1 + \lambda^2 \quad\quad\quad}{1 + \lambda^2 R_n^2/R_0^2}\right]^{1/4}_.
  \end{equation}

  This expression is tested in figure \ref{fig15B:rotor}a, where a contraction
  (resp. expansion) of the separation distance is observed for the wind turbine
  (resp. helicopter) regimes. While equation \eqref{eq29:model} gives correct
  trends and orders of magnitude, it tends to overestimate the expansion and
  underestimate the contraction. This difference could be explained in different
  ways. First, the contributions from the double-helix to $U_{t}$ estimated at
  the far-field from \eqref{eq:Hardinh} as $2R_0^* {d_0}/{h_{\tau,n}}$ have been
  neglected in \eqref{eq28:model} since vortex induction is rather weak. Second,
  a radial contraction of the double-helix close to the blade that should also
  be present but that has not been taken into account. Such contraction could be
  estimated by analogy to a vortex pair under a locally axial flow $U_t$.
  Considering these two effects, would give an additional contraction as
  observed in the numerics, but still would be too small to account for the
  discrepancy. Another possibility is the differential expansion/contraction
  experienced by the two vortices close to the rotor plane due to their radial
  position. The outermost vortex would experience a larger expansion/contraction
  than the innermost vortex, thus changing the separation distance. However,
  since the two vortices continuously trade places with one another, this
  effect is hard to quantify.

  In a similar vein, the pitch of the double-helix can be estimated using
  $h_{\tau} = 2\pi {U_{t}}/{\omega}$ where $\omega$ is the rotation rate of the
  vortex pair. As shown in figure \ref{fig3:approx}, for large values of
  $h_\tau/d$ we can approximate $\omega$ as $\omega = \Gamma/\pi d^2$.
  Combining both expressions shows that $h_{\tau,n}$ should be constant and
  given by
  \begin{equation}\label{eq30:model}
    \frac{h_{\tau,n}}{d_0}\approx \frac{2 \pi^2 }{\eta R_0^*} \left[\frac{\sqrt{1 + \lambda^2}}{\lambda}\right]
  \end{equation}
  which is verified in figure \ref{fig15B:rotor}b. Equation \eqref{eq30:model}
  also suggests that $h_{\tau,n}/d_0$ is primarily governed by $R_0^*$ and
  $\eta$, since the part shown in brackets remains close to 1 (in absolute
  value). The condition $h_{\tau,n}/d_0 > 10$ for being weakly twisted then
  reduces to $\eta R_0^* \lessapprox  2$.  For $\eta=0.04$, this condition is
  satisfied as soon as $R_0^*\lessapprox 50$.

  \begin{figure}
    (a) \hspace{0.25\linewidth} (b) \hspace{0.33\linewidth} (c)  \hfill ~ \\
    \includegraphics[width=\columnwidth, trim=0 0 0 0, clip]{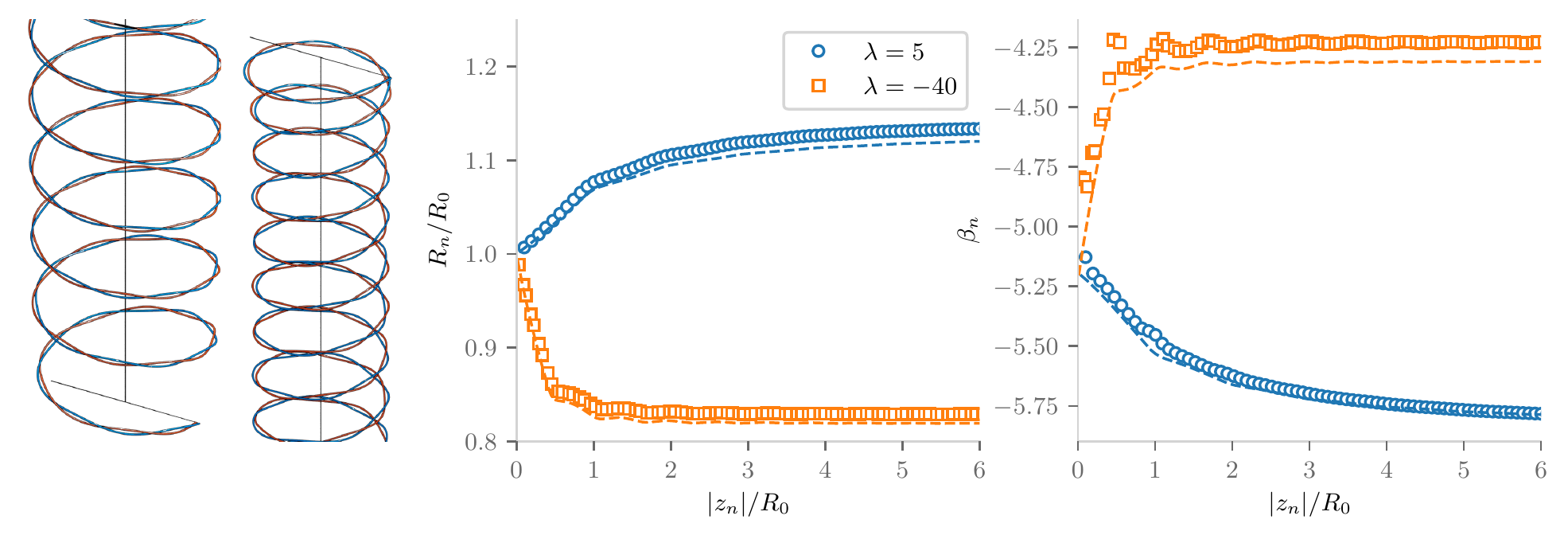}
    \caption{
    (a) Sparsely braided wakes obtained for $\eta = 0.04$, $R_0^* = 19.5$,
    $\varepsilon_0^{*} = 0.1$ in a wind turbine regime ($\lambda=5$) and a
    helicopter climbing flight regime ($\lambda=-40$). (b,c) Evolution of
    $R_n/R_0$ and $\beta_n$ compared to the equivalent helix approximation in
    dashed lines.
    }
    \label{fig18:rotor}
  \end{figure}

  From \eqref{eq30:model}, we can also compute the twist parameter
  $\beta_n=H_n/(h_{\tau,n} c_{\tau}(\zeta_n))$ as
  \begin{equation}\label{eq32:model}
    \beta_n \approx
   - \frac{\eta (R_0^*)^2}{\pi}
    \left[
    \frac{ |\lambda| }{ \sqrt{1 + \lambda^2 }}
    \frac{ \sqrt{H_n^2 + 4 \pi^2 R_n^2} }{2\pi R_0} \right].
  \end{equation}
  which has been compared to the data in figure \ref{fig15B:rotor}c. As the part
  shown in brackets remains in general close to 1, the typical order of
  $|\beta|$ is then given by the product $\eta (R_0^*)^2$. In particular, we
  expect to be in a leapfrogging situation, i.e. $|\beta|<1$, as soon as $\eta
  (R_0^*)^2 \lessapprox \pi$. Equation \eqref{eq32:model} suggests a simple way
  to control the twist parameter by increasing $R_0^*$ (i.e. reducing the
  separation distance $d_0$). In order to illustrate this point, we consider the
  same values of $(\lambda, \eta)$ for $R_0^*=19.5$ in different flight regimes
  and measure $\beta_n$. As expected, the evolution of the large-scale pattern
  is comparable to the equivalent single tip rotor, while $\beta_n$ is roughly
  four times larger with respect to the case with $R_0^*=9.5$ (compare figure
  \ref{fig15B:rotor}c and figure \ref{fig18:rotor}c).

  \begin{figure}
    \quad (a) \hspace{0.45\linewidth} \hspace{0.5cm} (b) \hfill ~ \\
    \includegraphics[width=\columnwidth, trim=0 155 0 5, clip]{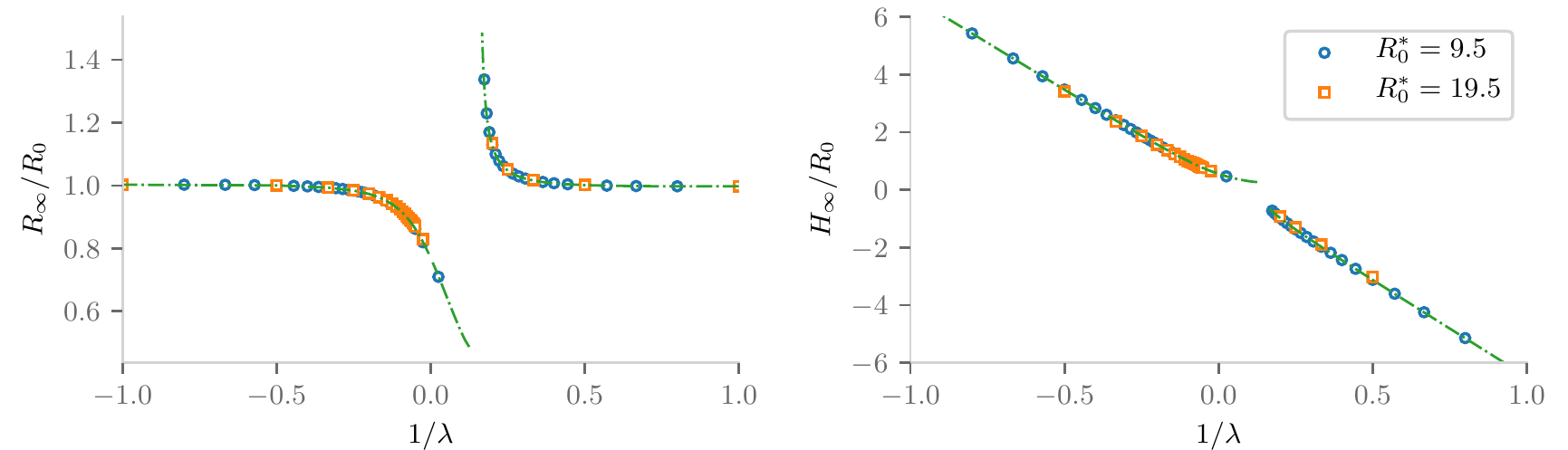}

    \quad (c) \hspace{0.45\linewidth} \hspace{0.5cm} (d)  \hfill ~ \\
    \includegraphics[width=\columnwidth, trim=0 5 0 155, clip]{./figure19_R1.pdf}
    \caption{
      Evolution of the far-wake parameters as a function of $1/\lambda$ for
      $\eta=0.04$, $\varepsilon _0^*=0.1$. Dashed lines correspond to the
      equivalent  vortex approximations in (a,b) and to equations
      \eqref{eq30:model} and \eqref{eq32:model} in (c,d). For reference, fig.
      (c) also displays the minimum and maximum values of $d(\zeta)/d_0$ for $\zeta\to\infty$ as a
      vertical bar. Helicopter regimes are on the left, wind turbine regimes on
      the right.
    }
    \label{fig16:rotor}
  \end{figure}
  \begin{figure}
    \quad (a) \hspace{0.45\linewidth} \hspace{0.5cm} (b) \hfill ~ \\
    \includegraphics[width=\columnwidth, trim=0 153 0 5, clip]{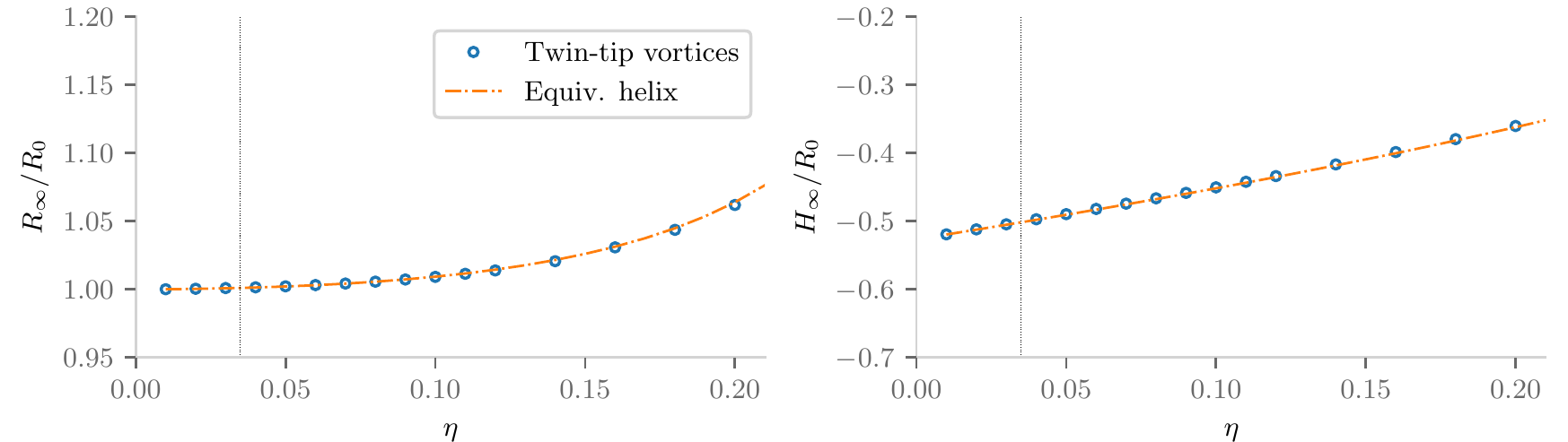}

    \quad (c) \hspace{0.45\linewidth} \hspace{0.5cm} (d)  \hfill ~ \\
    \includegraphics[width=\columnwidth, trim=0 7 0 155, clip]{./figure20_R1.pdf}

    \caption{
      Evolution of the far-wake parameters as a function of $\eta$ for
      $\lambda=2$ (wind turbine regime), $R_0^*=9.5$ and $\varepsilon _0^*=0.1$.
      Dashed lines correspond to the equivalent vortex approximations in (a,b)
      and to equations \eqref{eq30:model} and \eqref{eq32:model} in (c,d).  For
      reference,  fig. (c) also displays the minimum and maximum values of
      $d(\zeta)/d_0$ for $\zeta\to\infty$ as a vertical bar. The vertical dotted line in (c)
      indicates the value $\eta=\pi/(R_0^*)^2$ where $\beta_\infty \approx 1$.
      }
    \label{fig17:rotor}
  \end{figure}

  The above approximations can also be used to predict the characteristics of
  the far-wake. In figures \ref{fig16:rotor} and \ref{fig17:rotor}, we have
  compared the model with the numerics as both $\lambda$ and $\eta$ vary. We can
  observe in figures \ref{fig16:rotor}a,b and \ref{fig17:rotor}a,b  that for the
  radius $R_\infty$ and the pitch $H_\infty$, the agreement with the equivalent
  vortex approximation is excellent for all regimes. For the separation distance
  $d_\infty$ (figures  \ref{fig16:rotor}c and \ref{fig17:rotor}c), the trend is
  qualitatively good but there is a constant systematic over-estimation in the
  model, that has already been noticed above. The fact that $d_\infty$ changes
  little with $R_0^*$ is also consistent with \eqref{eq30:model}. Interestingly,
  the twist parameter $\beta_\infty$ (figures \ref{fig16:rotor}d and
  \ref{fig17:rotor}d) is always found to be well-predicted, probably because it
  does not directly depend on the separation distance.


  \section{Conclusions}

  In this article, we have provided a vortex wake model for a tip-splitting
  rotor in any incident normal wind, which applies to all the vertical
  helicopter flight and wind turbine regimes. This model is based on a
  simplified description of the wake with a pair of co-rotating
  vortex filaments attached to each blade at different radial positions. A
  free-vortex method together with Biot \& Savart law has been used to obtain
  steady solutions in a frame moving with the rotor blades.

  We have first analyzed the structure of the solutions in the far-field. As for
  counter-rotating vortices \cite{venegas2019generalized}, we have shown that
  steady periodic solutions can be obtained in specific reference frames and that the
  properties of these solutions are governed by 4 purely geometrical
  dimensionless parameters ($R/d$, $H/d$, $h_\tau/d$ and $a/d$). We have
  documented the fluctuations of the separation distance and of the mean radius,
  as well as the variations of the frame velocities  with respect to two of
  these parameters ($H/d$ and $h_\tau/d$). We have shown that these solutions
  can be understood in terms of two interlaced helical vortices inscribed on top
  of a larger helical structure, which allowed us to obtain good estimates
  for the wake geometry and the frame velocities.

  The far-field solutions have then been used to get near-wake solutions. We have
  solved the semi-infinite free-vortex problem with a prescribed condition on
  the blade (the vortices are attached at given positions) and a condition of
  matching with a far-field solution away from the rotor. We have been able to
  obtain steady solutions in the frame rotating with the rotor, that capture the
  expected global expansion/contraction of the wake as well as the
  deformations induced by the mutual interaction of the vortices. Interestingly,
  we have also been able to show that the solution is well described by a
  small-scale twisted structure inscribed onto a large-scale pattern obtained for
  a single large tip-vortex with the total circulation. We have further shown
  how the variations of the parameters of the small-scale structure are driven
  by the large-scale motion.


  In applications, the core size of tip vortices is sufficiently small to make
  the filament approach relevant. For instance, in the MEXICO project, a wind
  turbine of diameter 4.5 m creates a tip vortex of radius 2 cm giving a ratio
  $a/R \sim 0.01$ \cite{snel2007mexico}. Similar values can be found in
  helicopter rotor wakes \cite{bauknecht2017blade}. In such cases, there is
  clearly enough room to generate a second tip vortex at a small distance $d$
  from the tip such that $a/d$ remains small. We therefore claim that the
  typical parameters $a/d=0.1$ and $R/d=10$ that we have considered in the
  present study  can be obtained in real applications. However,  it does not
  mean that our analysis will automatically apply. The roll-up process giving
  rise to the vortices could be significantly more complex. In a recent
  experimental study, \citet{Schroder2021} succeeded in creating a second
  vortex close to the tip using a fin. However, the fin vortex was also found to
  contain negative vorticity contributions coming from the vorticity sheet shed
  by the modified blade, that made it unstable with respect
  to the centrifugal instability. In that case, the second vortex is thus
  rapidly disrupted and our solution cannot be formed. We suspect that if the
  blade was more tapered close to the tip, the fin vortex would not have been
  unstable and a different evolution would have been observed.

 Our solutions are nevertheless expected to be unstable. Helical solutions are
 indeed known to be unstable with respect to a long-wavelength instability, the
 so-called Widnall instability
 \citep{widnall1972stability,gupta1974theoretical}. As shown in
 \citet{quaranta2019local} and \citet{huang2019numerical}, this instability can be excited
 to accelerate the destruction of the vortices.
 In a follow-up paper \citep{Castillo2020b}, the linear stability of our
 solutions is analyzed using the filament framework.
 Owing to the higher complexity of our solutions, we demonstrate that there
 exist different types of instability modes. As for the Widnall instability
 \citep{quaranta2015long,quaranta2019local}, each instability mode is
 nevertheless shown to be associated with a specific pairing event.

Short-wavelength instabilities such as the curvature instability
\cite{blanco2017curvature} and the elliptic instability
\cite{kerswell2002elliptical,blanco2016elliptic} are also expected to develop in
the core of the vortices. These instabilities cannot be described  by the
filament approach which  neglects everything occurring in the vortex cores. It
requires monitoring the core deformations induced by curvature, torsion and
strain fields which are responsible for these instabilities. This can be done by
matched asymptotic techniques  \cite{callegari1978motion,blanco2015internal} or
by direct numerical simulations when the base flow exhibits particular
symmetries as shown by \cite{ledizes2002viscous} for 2D vortex pairs,
\cite{hattori2019numerical} for rings and \cite{Selcuk_2017} for helices. The
internal core structure that is obtained can then be used for the
short-wavelength stability study (see for instance
\cite{roy2008stability,hattori2019numerical}).

Note that the determination of a quasi-steady solution may not always be
necessary for the stability study  if the time scale of the instability is long
compared to the relaxation time scale needed to get the correct core structure.
\citet{brynjell2020numerical} have indeed shown that they could obtain the
correct stability properties of a helix with a Gaussian vorticity core by just
analysing the temporal evolution of the numerical solution obtained from the
initial condition formed by the filament structure with the prescribed
axisymmetric vortex profile in the core. This is an interesting result as it
provides a simple way to analyse the short-wavelength stability properties of
our solutions.

  \section*{Acknowledgment}

  This work is part of the French-German project TWIN-HELIX, supported by the
  Agence Nationale de la Recherche (grant no. ANR-17-CE06-0018) and the Deutsche
  Forschungsgemeinschaft (grant no. 391677260).

  \bibliography{articles}

\end{document}